\RequirePackage{fix-cm}

\documentclass[smallextended]{svjour3}       % onecolumn (second format)
\smartqed  % flush right qed marks, e.g. at end of proof
\usepackage{graphicx}
\usepackage{multirow}
\usepackage{multicol}
\usepackage[a4paper, margin=2cm]{geometry}
\begin{document}

\newcommand{\procspie}{\rm{Proc.~SPIE}}
\newcommand{\apjs}{\rm{ApJS}}
\newcommand{\apj}{\rm{ApJ}}
\newcommand{\aap}{\rm{A\&A}}
\newcommand{\pasp}{\rm{PASP}}
\newcommand{\nat}{\rm{Nature}}
\newcommand{\ao}{\rm{Applied Optics}}

\title{TARdYS: Design and Prototype of an Exoplanet Hunter for TAO using a R6 Echelle Grating}

%\subtitle{Do you have a subtitle?\\ If so, write it here}

\titlerunning{TARdYS: Design and Prototype}        % if too long for running head

\journalname{Experimental Astronomy}

\author{S.~Rukdee$^{1,2,*}$\and L.~Vanzi$^{1,2}$\and C.~Schwab$^{3,4}$\and M.~Flores$^{1,2}$\and A.~Valenzuela $^{1,5}$\and J.~Pember$^{3}$\and A.~Zapata$^{1,2}$\and K.~Motohara$^{6}$\and Y.~Yoshii$^{6,8}$\and M.~Tala Pinto$^{7}$ }

\institute{{*}
\email{surangkhanar@gmail.com} \\
{1} Center of Astro Engineering UC (AIUC), Av. Vicuna Mackenna 4860, Macul - Santiago, Chile\\ 
{2} Department of Electrical Engineering, Pontificia Universidad Catolica de Chile, Av. Vicuna Mackenna 4860, Santiago Chile\\
{3} Macquarie University, Balaclava Road, North Ryde NSW, 2109, Australia\\ 
{4} Australian Astronomical Observatory Sydney, Australia\\ 
{5} Institute of Astrophysics, Pontificia Universidad Catolica de Chile, Av. Vicuna Mackenna 4860, Santiago Chile\\ 
{6} Institute of Astronomy, Graduate School of Science, the University of Tokyo, 2-21-1 Osawa, Mitaka, Tokyo 181-0015, Japan\\ 
{7} Landessternwarte, Zentrum fur Astronomie der Universitat Heidelberg, Konigstuhl 12, D-69117, Germany \\
{8} Steward Observatory, University of Arizona, 933 North Cherry Avenue, Rm. N204, Tucson, AZ 85721-0065, USA }

\authorrunning{S. Rukdee et al.} % if too long for running head

\date{Received: date / Accepted: date}
% The correct dates will be entered by the editor

\maketitle

\begin{abstract}
One limitation in characterizing exoplanet candidates is the availability of infrared, high-resolution spectrographs. An important factor in the scarcity of high precision IR spectrographs is the high cost of these instruments.  We present a new optical design, which leads to a cost-effective solution. Our instrument is a high-resolution (R=60,000) infrared spectrograph with a R6 Echelle grating and an image slicer. We compare the best possible performance of quasi-Littrow and White Pupil setups, and prefer the latter because it achieves higher image quality. The instrument is proposed for the University of Tokyo Atacama Observatory (TAO) 6.5 m telescope in Chile. The Tao Aiuc high Resolution (d) Y band Spectrograph (TARdYS) covers 0.843-1.117 $\mu$m. To reduce the cost, we squeeze 42 spectral orders onto a 1K detector with a semi-cryogenic solution. We obtain excellent resolution even when taking realistic manufacturing and alignment tolerances as well as thermal variations into account. In this paper, we  present early results from the prototype of this spectrograph at ambient temperature.
\keywords{NIR Spectroscopy \and Instrumentation \and Radial Velocity \and Exoplanet}
\end{abstract}

\section{Introduction}

After the first exoplanet discovery in 1995 \cite{1995Natur.378..355M}, astronomers have been searching and characterizing properties of exoplanets in the visible wavelength region using various methods: Radial velocity, Astrometry, Timing, Microlensing, Transits, and Direct imaging \cite{2011exha.book.....P}. The Radial Velocity (RV) technique measures Doppler shifts in stellar lines due to motion produced by a nearby planet using a high-resolution spectrograph. Despite an increasing number of exoplanet candidates found with the transit technique (\cite{2016ApJS..226....7C}, \cite{2016ApJS..222...14V}), RV measurements are a fundamental and reliable tool for confirming and characterizing exoplanets. 

Recently, the search has moved to cooler and lower mass stars. Due to their large numbers, M dwarfs are prime candidates for planets in their habitable zone, a distance from a star where water stays liquid \cite{2007AsBio...7...30T}. Infrared (IR) spectroscopy is highly efficient to observe M type stars, since the IR flux at 1 $\mu$m (Y-band) is several times higher than in the optical regime. To achieve RV precision of 2.2 m/s in Y-band spectroscopy, assuming signal to noise (S/N) of 100 at 1 $\mu$m, requires spectral resolution R=60,000 for spectral-type M9 \cite{2010ApJ...710..432R}.

In recent years, several near-infrared spectrographs\cite{2014Natur.513..358P} started operation, such as CRIRES \cite{Moorwood2005}, GIANO \cite{2006SPIE.6269E..19O}, IGRINS \cite{2010SPIE.7735E..1MY}, HPF \cite{2014SPIE.9147E..1GM}, IRD \cite{2012SPIE.8446E..1TT}, CARMENES \cite{2014SPIE.9147E..1FQ}, SPIRou \cite{2014SPIE.9147E..15A}, and WINERED \cite{2016SPIE.9908E..5ZI}. In the Table \ref{tab:spec_compare}, we compare some important parameters for these recent spectrographs optimized for RV measurements. 

\begin{table}[h!]
\caption {Design comparison of existing fiber-fed near-infrared spectrographs optimized to search for exoplanets.} 
\label{tab:spec_compare}
\begin{center}
\begin{tabular}{ lcclclc }
\hline
Instrument & $\lambda$ [$\mu$m] & R & Telescope & Config. & Input & Main Disperser\\
\hline
CARMENES &	0.90-1.75 &	80,000 &	Calar Alto 4m  &	WP &	Fiber-IS* &	Echelle R4 \\ 
IRD	& 0.97-1.75 & 	70,000 &	Subaru 8m  &	WP &	Fiber &	Echelle R6 \\ 
HPF &	0.85-1.70 &	60,000 &	HET 10m	 &	WP &	Fiber-IS &	Echelle R4 \\ 
SPIRou &	0.90-2.40 &	85,000 &	CFHT	3.6m &	WP &	Fiber &	Echelle R4 \\ 
\textbf{TARdYS} &	\textbf{0.84-1.12} &	\textbf{66,000} &	\textbf{TAO 6.5m}  &	\textbf{WP} &	\textbf{Fiber-IS} &	\textbf{Echelle R6} \\ 
\hline
\end{tabular}
\end{center}
* WP = White Pupil, Fiber-IS = Fiber and image-slicer
\end{table}

From Table \ref{tab:spec_compare}, we see that the number of spectrographs is limited. Furthermore, the existing spectrographs are all installed at the telescopes in the northern hemisphere, and there are not enough instruments available compared to the number of targets. This is because building one high resolution instrument with an appropriate stability and precision for the RV measurement is usually very expensive. In this work, we develop a design to lower the cost and make the spectrograph more compact. 

TARdYS (The Tao Aiuc high Resolution (d) Y band Spectrograph) is a unique project designed to test less common components. These include a R6 Echelle grating, the image slicer in the near-infrared region, only a semi-cryogenics setup, and working at a very high altitude (5,640 m) environment at the University of Tokyo Atacama Observatory (TAO: Project PI Yuzuru Yoshii), the highest observatory site in the world \cite{doi:10.1117/12.2231391}. 

This paper presents the optical design study and prototype performance of TARdYS. We explain our optical design in section 2 and provide an analysis in section 3, which includes tolerance analysis, thermal analysis and the exposure time calculator of the expected signal to noise. We have built a prototype and in section 4 present optomechanical design and construction. Finally we evaluate the radial velocity drift and temperature stability of the prototype.

\section{Optical Design}
The optical design considerations for a near-infrared spectrograph are similar to those working in the visible wavelength range except for the optical material, cost and the IR detector. High thermal background is a major problem for designing near-infrared spectrographs. As a result, these spectrographs are normally cooled in a cryogenic system.

Our spectrograph is designed according to the science requirements and the budget constraint of having a IR detector not larger than 1,024 x 1,024 pixels size. To achieve R $\geq$ 50,000 with a 1k detector requires operating at order m $>$ 100. This leads to the main disperser choice of a commercially available Echelle R6 grating with 13.33 grooves/mm giving a blaze angle = 80.6$^{\circ}$. It disperses all the Y-band within the detector's limited space. The choice of the 1K detector and a R6 grating set the effective focal length (EFL) of the objective to 200 mm and yield a 2 pixel Nyquist sampling.

We use a fiber at our spectrograph entrance aperture for a good mechanical stability of the instrument. The light from dual fibers (50 $ \mu$m diameter, f/4), which corresponds to 0.4 arcsecond diameter on the sky, is fed through a simplified version of the Bowen-Walraven image slicer (\cite{1938ApJ....88..113B}, \cite{doi:10.1117/12.856709} and \cite{2017ExA....43..167T}). 

\subsection{Setup Configuration}
%\end{figure}

The design study considers two different configurations: quasi-Littrow and White Pupil.

\subsubsection{quasi-Littrow Configuration}
\label{sec:QL}
\begin{figure}[h!]
A) quasi-Littrow configuration:\\
\includegraphics[width=0.99\columnwidth]{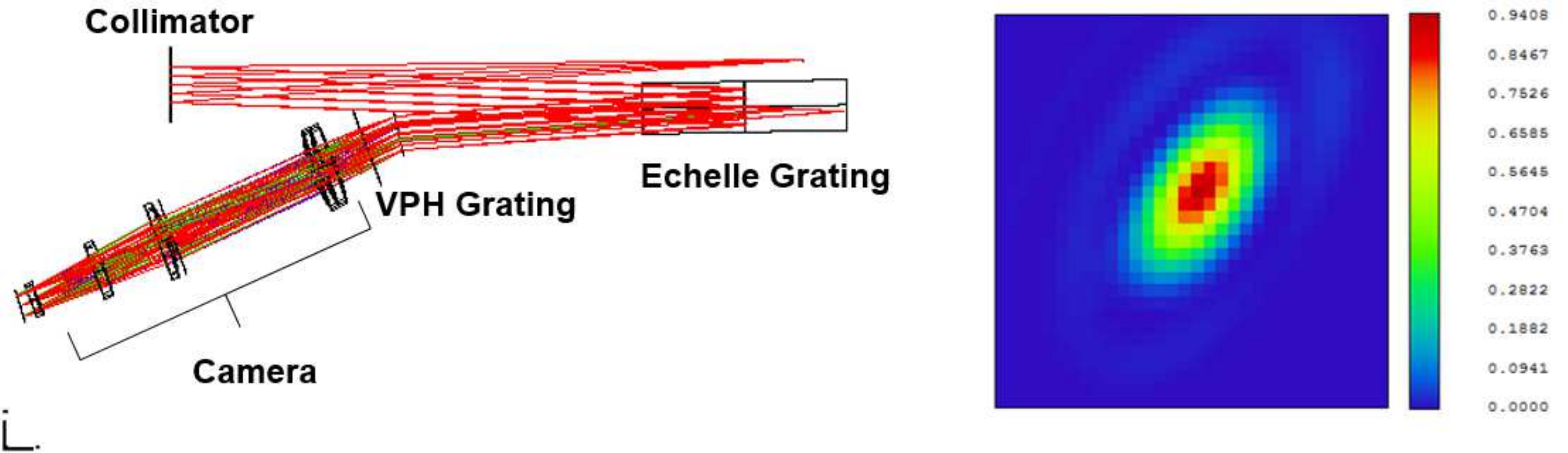}\\

B) White Pupil configuration:\\
\includegraphics[width=0.99\columnwidth]{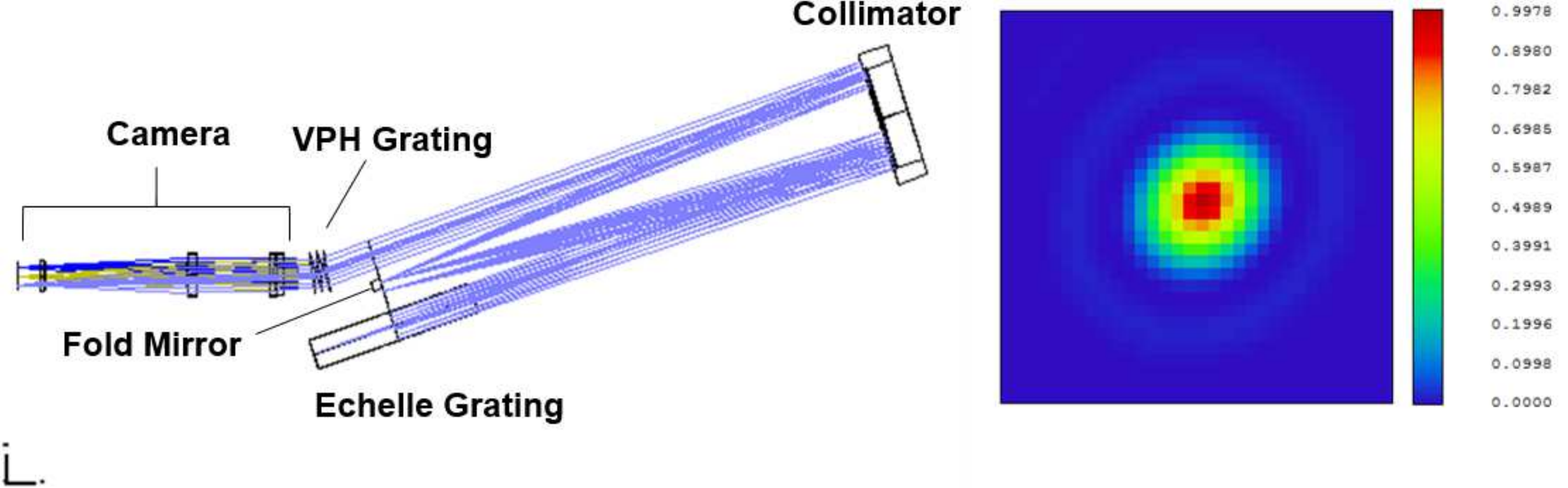}
\caption{Optical layout of quasi-Littrow configuration (top left panel) and White Pupil configuration (bottom panel). The corresponding plots of the point spread function (PSF) on the right shows a symmetric behavior, due to the White Pupil configuration, on the image plane. The strehl ratio calculated at central wavelength or $\approx 1 \mu$m of this configuration reach 0.99 in the red region.}
\label{fig:Fig1A}
\label{fig:Fig1B}
\end{figure}

In the concept design of TARdYS, we started with a quasi-Littrow (QL) spectrograph design \cite{2012SPIE.8446E..81B}. As shown in Fig. \ref{fig:Fig1A}, the light beam reflects on a collimating mirror, an Echelle grating, a VPH grating and finally the camera optics respectively. With this configuration, we found that the predicted performance of the spectrograph from ray-tracing suffers from anamorphism. The focal plane image is always tilted due to the setup of the spectrograph as shown in Fig. \ref{fig:Fig1A}. 

\subsubsection{White Pupil Configuration}
\label{sec:WP}
We apply Baranne's classic White Pupil (WP) configuration \cite{1972ailt.conf..227B} and a parabolic mirror (conic constant K = -1) as a collimator in Fig.  \ref{fig:Fig1B}. Gratton's analytic calculations of Third-Order aberration \cite{2000ApOpt..39.2614G} demonstrate that the WP configuration can cancel spherical aberration, coma, astigmatism, distortion and chromatic aberrations. The White Pupil configuration re-images a pupil placed on a grating into a secondary pupil where a cross disperser or camera optics can be placed.  It is called 'white' because the pupil image location is independent of wavelength even though the light is dispersed \cite{2009arXiv0910.0167B}. It minimizes the geometrical aberrations that occur in QL configurations. The design delivers better image quality than the QL design for the same resolution. In the right panels of Fig. \ref{fig:Fig1A}, we evaluate the point spread function (PSF) to compare the image quality of the different design options.  Both yield a similar value of strehl ratio in central field shown in red in Fig. \ref{fig:Fig1A} at approximately 1.0 $\mu$m. However, the QL configuration shows asymmetrical and tilted PSF behavior. In contrast, the WP PSF is circular.\\

\subsection{Dispersers}

\begin{figure}[h!]
\center
\includegraphics[scale=0.2]{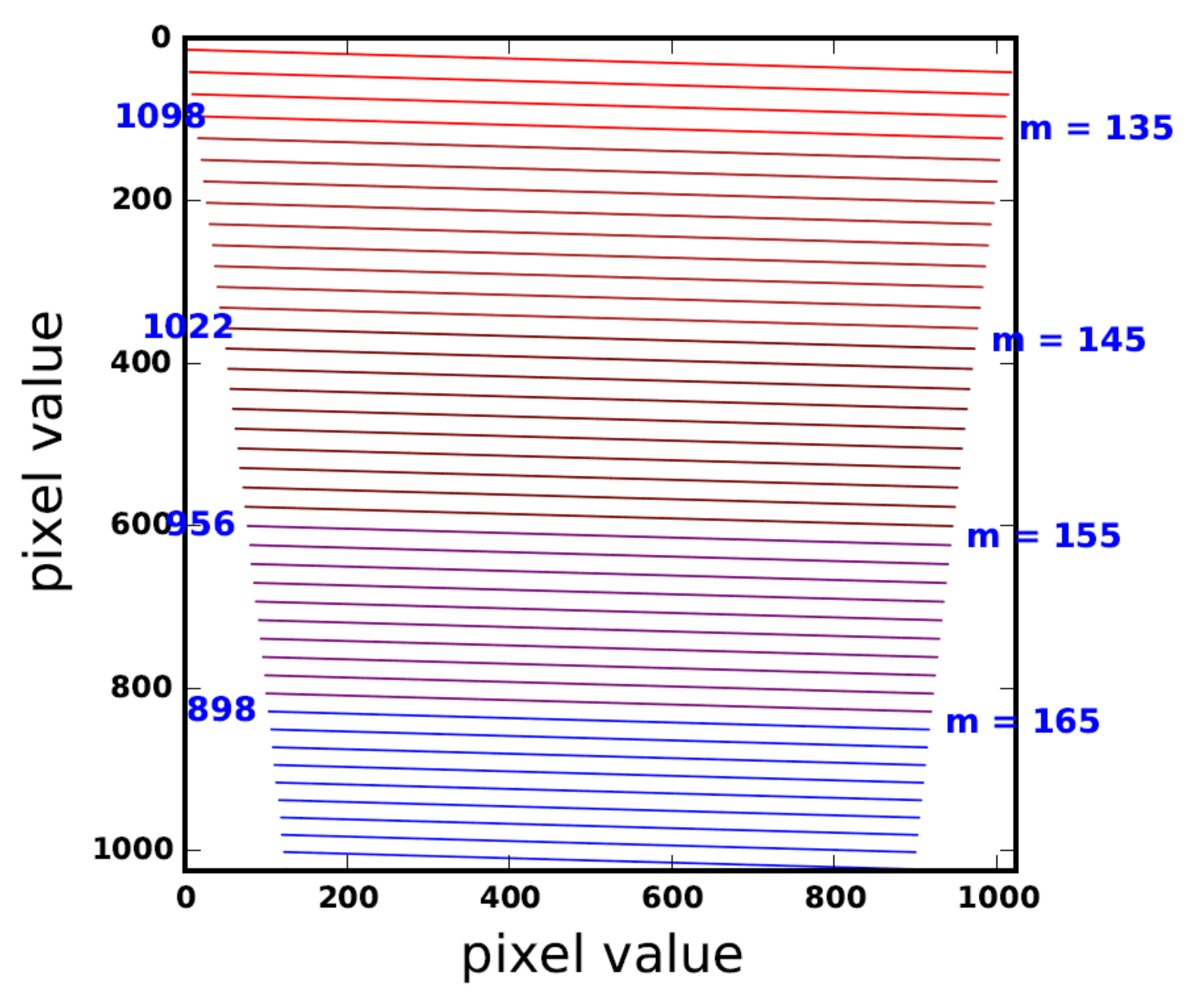}
\includegraphics[scale=0.2]{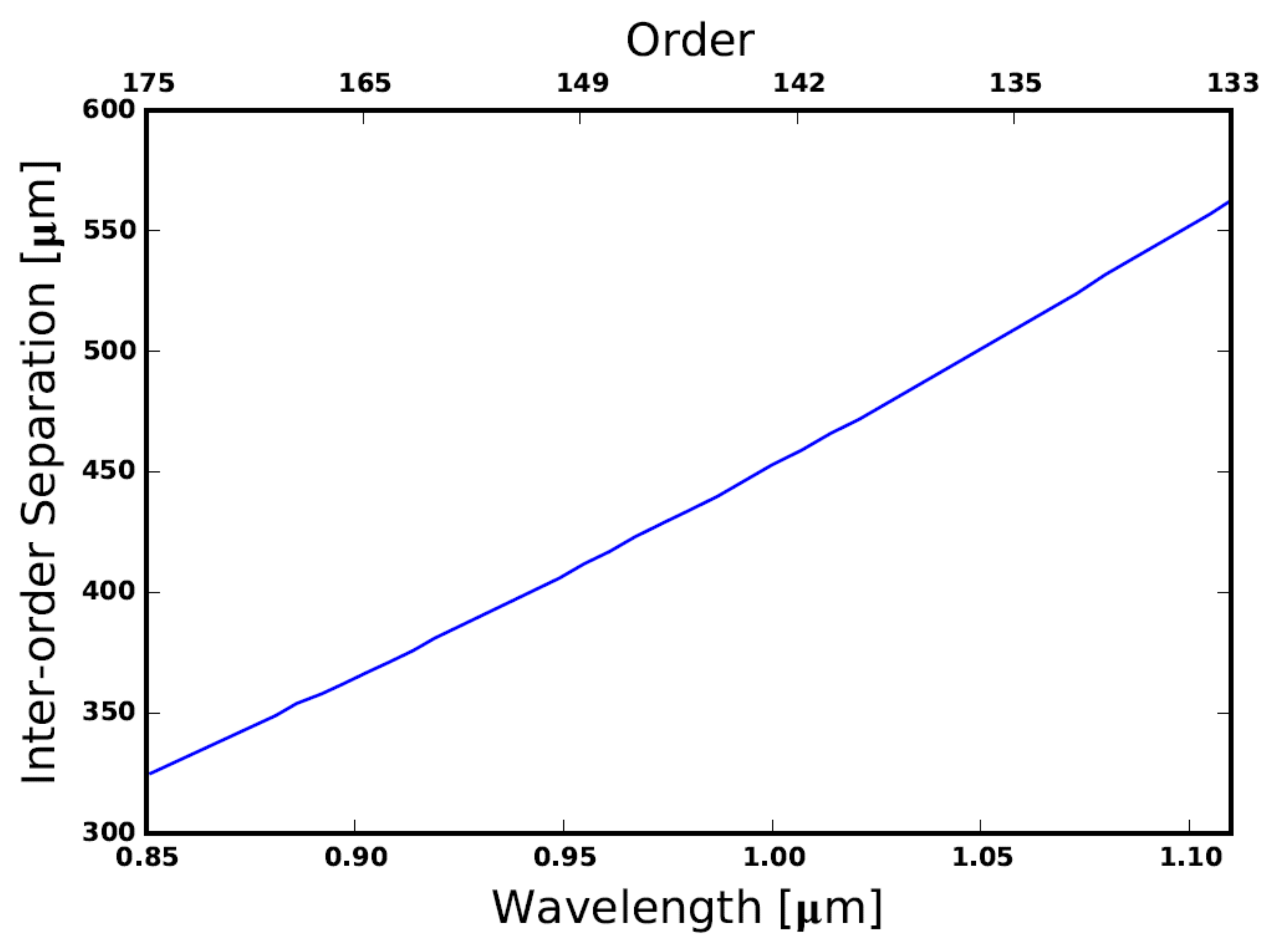}
\caption{The echellogram in the left panel shows the free spectral range of 42 orders on the detector of 1,024x1,024 pixels. The right panel shows inter-order separation using VPH grating with 333 lines/mm as cross-disperser.}
\label{fig:echellogram}
\end{figure}

Our main dispersive element is a R6 Echelle grating. Echelle gratings are blazed to high angles to yield high orders but this causes order overlapping that must be separated by cross-dispersion. The R values correspond to the tangent of a grating's blazed angle. The difference of these angles also has an effect on the size and volume of the spectrograph; smaller R value corresponds to larger spectrographs. In order to have a compact spectrograph, we choose a R6 Echelle grating, which gives the largest blazed angle of 80.5 degrees commercially available to date. It is the only fixed component in our design. The echellogram shown in Fig. \ref{fig:echellogram} is calculated based on a diffraction equation from Schroeder \cite{1967ApOpt...6.1976S}. We apply $f_{col}$ = 550 mm. And our spectrograph will achieve R $\approx$ 66,000.

We choose a Volume Phase Holographic (VPH) Transmission Grating from Kaiser Optical System, Inc. as a cross disperser to obtain high efficiency. To optimize the inter-order separation, we modeled the dispersion in ZEMAX and tested various options of VPH groove spacing. The separation increases with wavelength to the redder region. This is shown in the right panel of Fig. \ref{fig:echellogram}. We decided to use a VPH grating with 333 lines/mm at a 9.3$^\circ$ incidence angle, which is ideal because the two gratings will disperse the incoming light over a 0.843-1.117 $\mu$m  wavelength range with a total of 42 orders (133 to 175) on our 1k detector. We simulate the detector plan according to the parameters mentioned above. The echellogram on the detector is shown in the top left panel of Fig.\ref{fig:echellogram}. 

\subsection{Camera}

Our goal for optimizing the camera is to have the least number of surfaces in order to minimize losses. We design the camera of TARdYS for room-temperature assembly to minimize costs. Nevertheless, to achieve low noise, the detector and the last lens are placed in a cryogenic Dewar cooled to 80K. The purpose of having this semi-cryogenic setup is to study the temperature effect of a Y band spectrograph later.  
\begin{figure}[h]
\centering\includegraphics[width=0.7\textwidth]{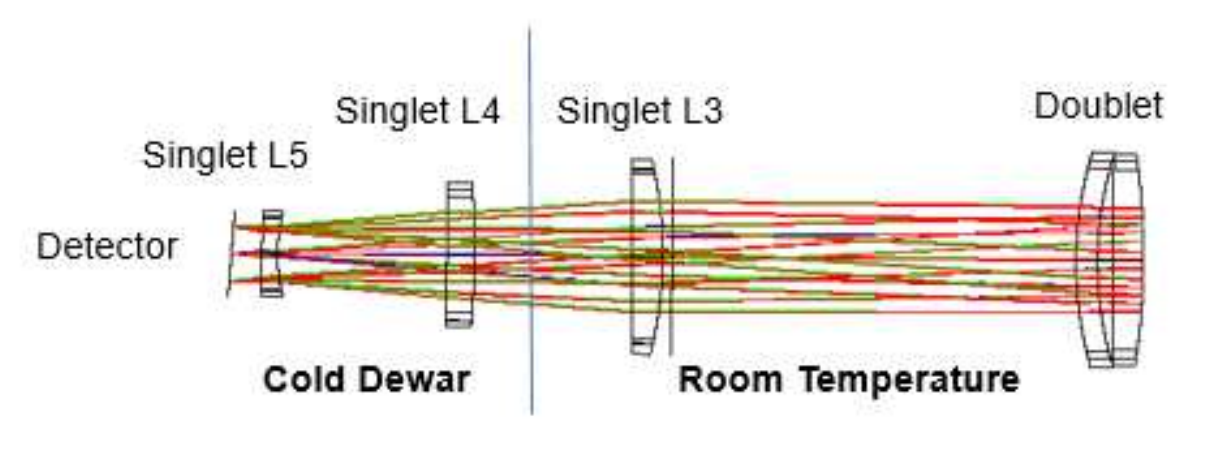}
%\caption{}
\centering\includegraphics[width=\textwidth]{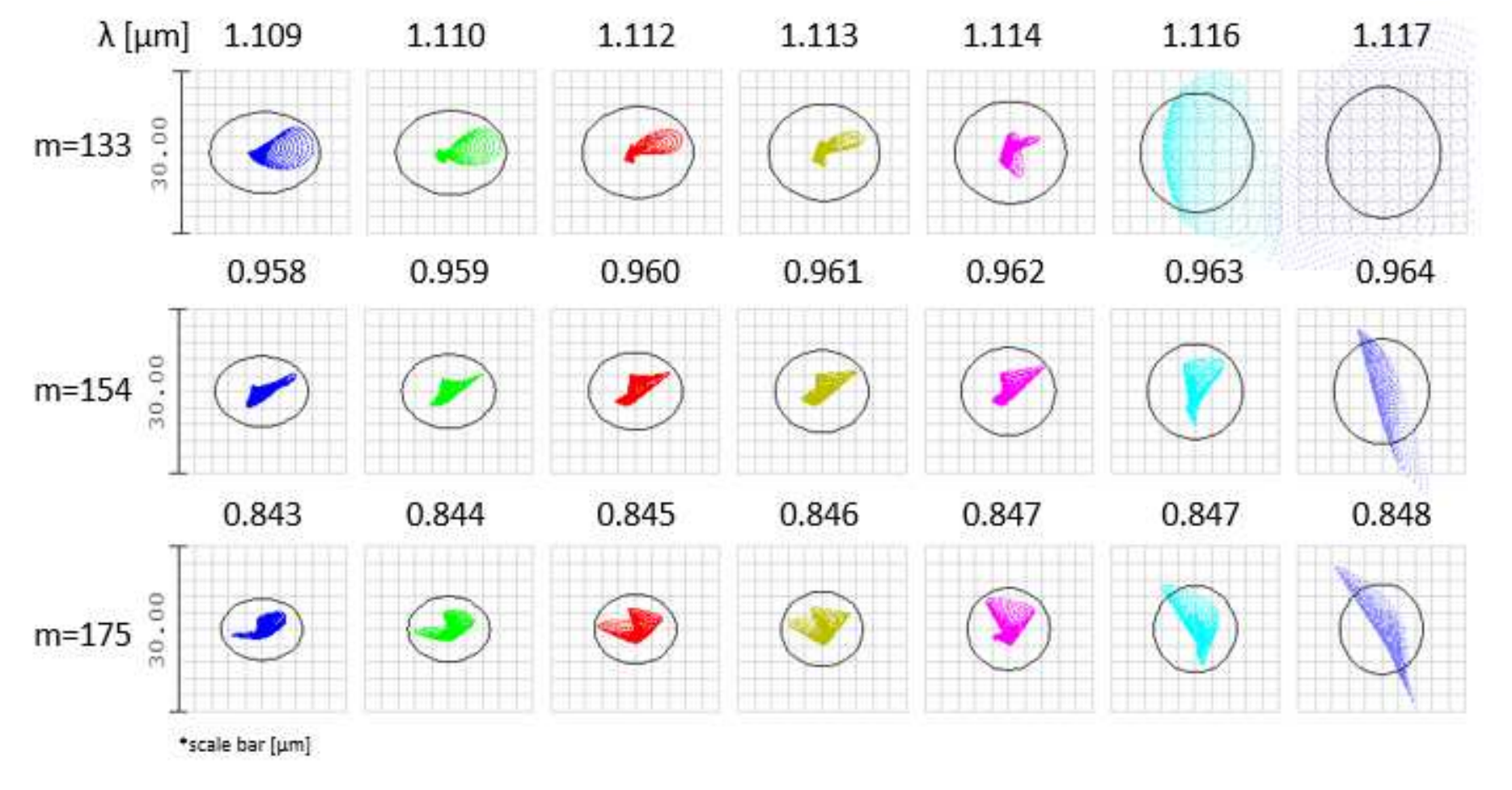}
\caption{Top panel: the camera design with 5 lenses. The spot diagrams in the lower panel show the rays traced across the focal plane on the image surface from the 5 lenses camera. The three rows present the reddest, central and bluest orders. Columns present increasing wavelength within each order (m).}
\label{fig:camera1}
\end{figure}

\begin{figure}[h]
\centering\includegraphics[width=0.7\textwidth]{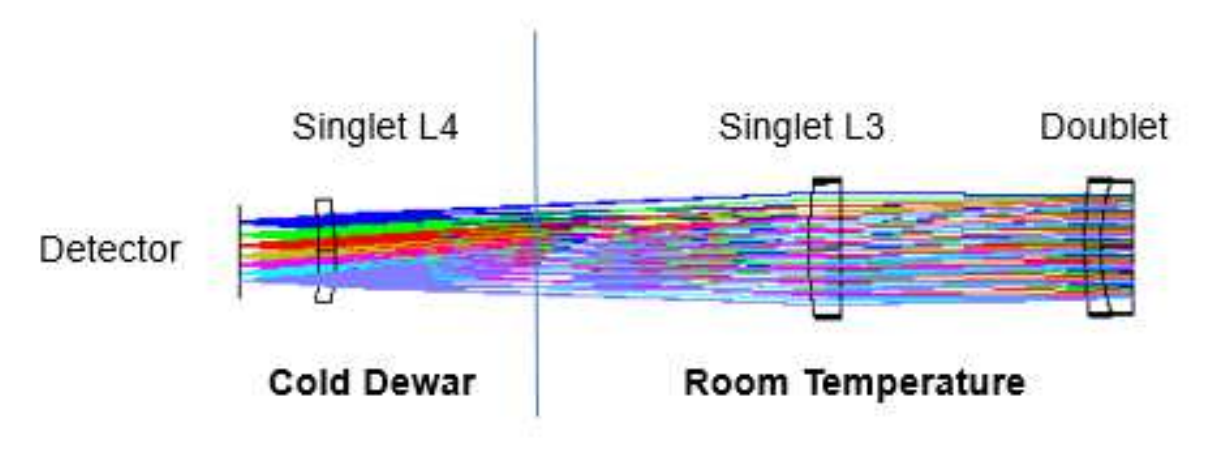}
\centering\includegraphics[width=\textwidth]{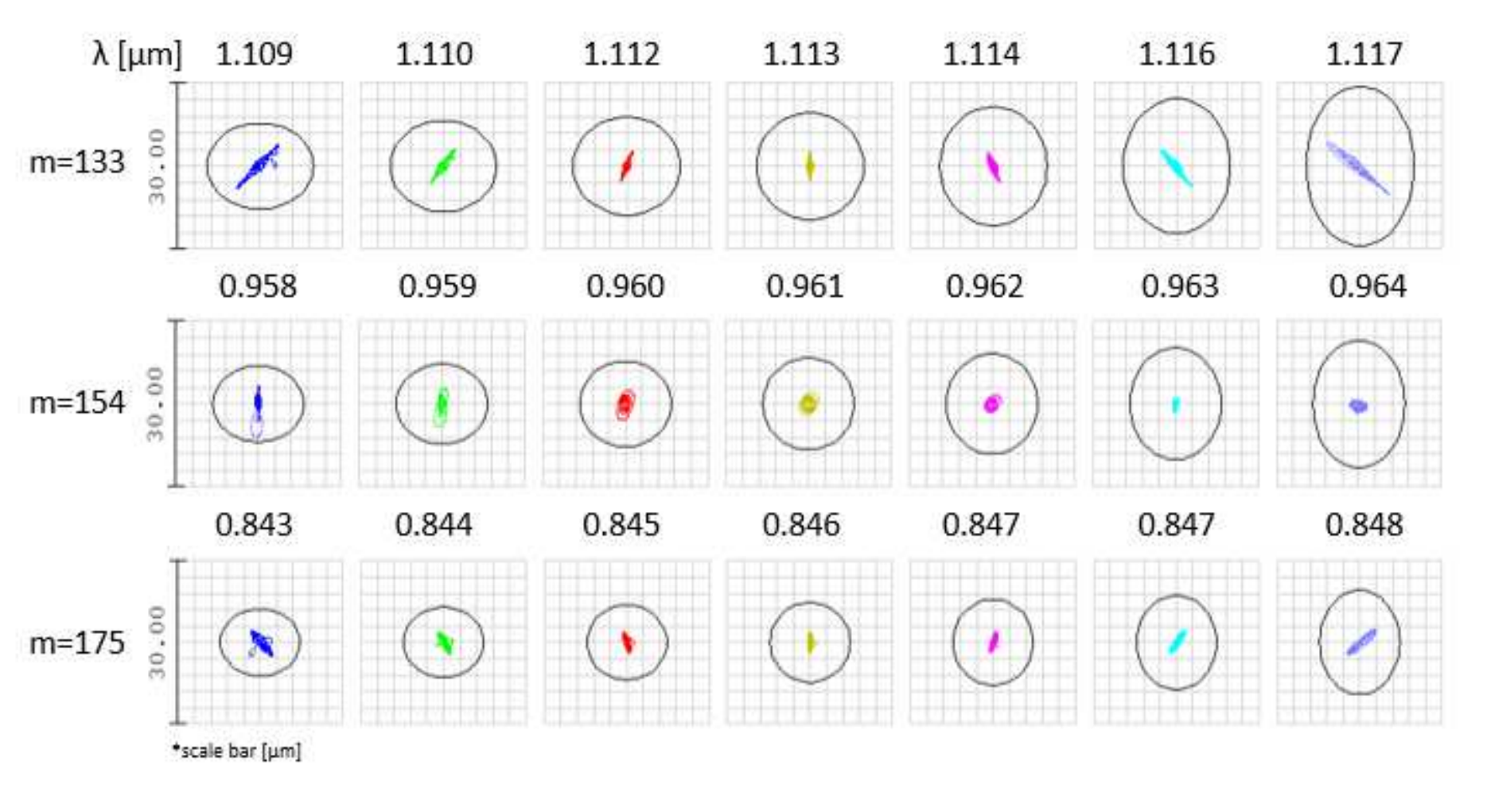}
\caption{Final camera design with 4 lenses. The last (left) part of this camera, the detector and the last singlet L4, will be cooled in cryogenic Dewar. Spot diagrams below show the rays traced across the focal plane on the image surface, as in Fig.\ref{fig:camera1}. In all cases, the spots are smaller than the diffraction limit (black circles). The three rows present the reddest, central and bluest orders. Columns present increasing wavelength within each order.}
\label{fig:camera2}
\end{figure}

We choose a doublet lens at the beginning of the camera  to correct for chromatic aberration. The design of the camera started with 5 lenses: a doublet and 3 singlets. The choice of doublet is a simple combination of N-BAK4 and N-SF10. The singlets are BK7 glass, which has a relatively low cost. We show the camera lenses in the top panel of Fig. \ref{fig:camera1}. We optimized the camera to obtain small spot sizes within the diffraction limit. The bottom panel in Fig. \ref{fig:camera1} shows the spot matrix for the lowest, middle and highest orders. Generally, the spot sizes increase from left to right, where it exceeds the diffraction limit of the airy disk (black circle). In addition, the spot sizes shown in Fig. \ref{fig:camera1} increases rapidly at one side of the detector and appear to be defocused at the upper right corner.

We evolved the camera design from this point by first changing the doublet combination. A crown glass and a flint glass were selected by the hammer optimization tool in ZEMAX considering the transmission wavelength and relative cost. We then decreased the number of singlets by carefully optimizing the ZEMAX merit function and flattening one middle lens until it could be removed. 

Our final design is depicted in Fig. \ref{fig:camera2}. We select one doublet of N-SK2 and N-SF6. The two singlets (L3 and L4) are fused-silica lenses. These are less affected by thermal expansion because we plan to cool this part of our camera. In the end, the camera has a focal length of 200 mm and consists of four 50 mm diameter lenses. Only the last lens will be in the cryogenic Dewar. Spot diagrams of rays traced across the focal plane on the image surface. In Fig. \ref{fig:camera2}, the RMS spot size is clearly smaller than the 5 lenses configuration in Fig. \ref{fig:camera1}. It is also diffraction limited everywhere in the ambient environment. 

%\newpage
\subsection{Comparison}
\begin{figure}[h!]
\centering\includegraphics[scale=0.3]{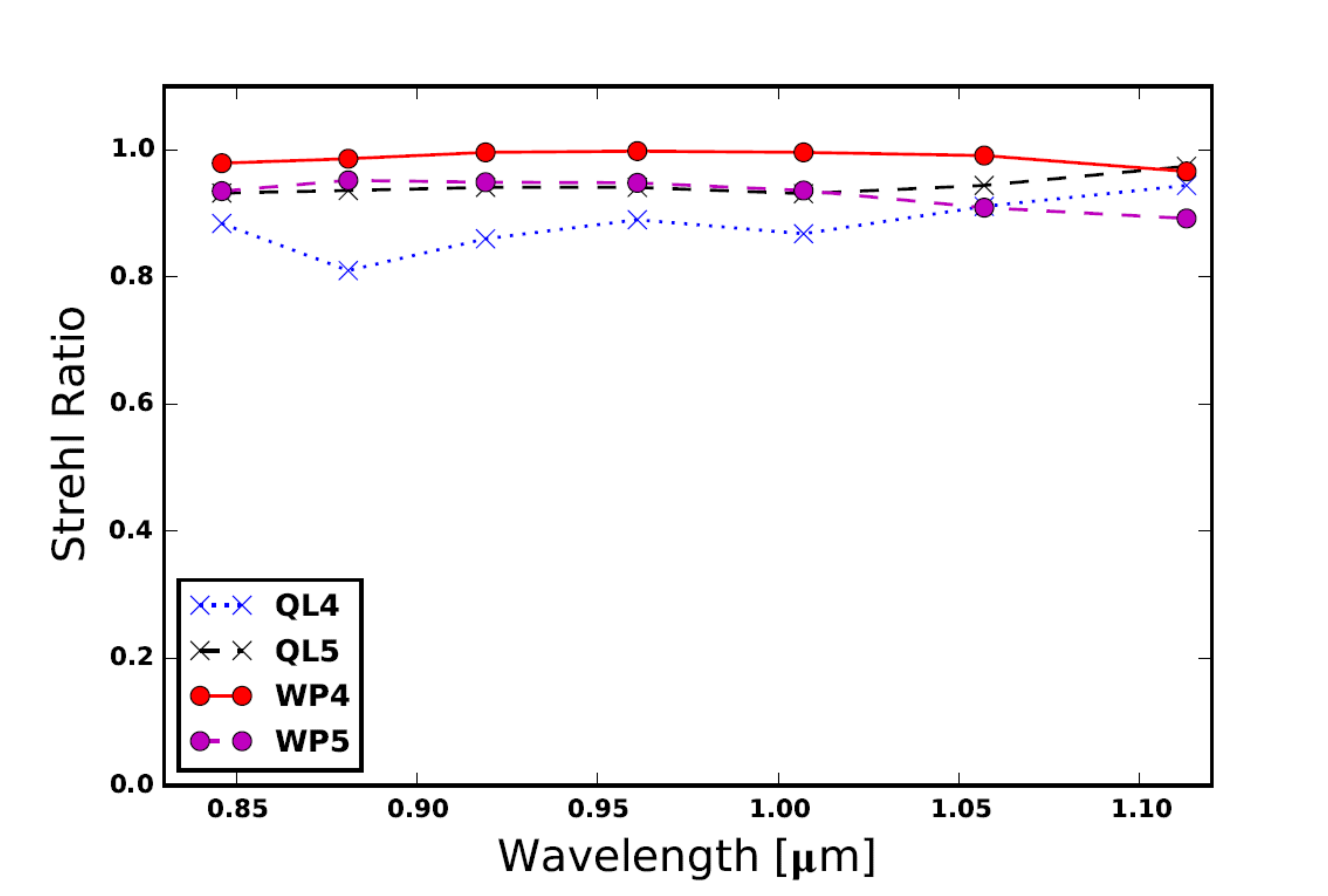}
\caption{Strehl ratio of four configurations: quasi-Littrow (QL) or White-pupil (WP) with 4 or 5-lens cameras. The White-pupil configuration with 4-lens camera (red line) yields the highest ratio throughout the working wavelength range.}
\label{fig:Strehl_ratio}
\end{figure}

We now compare the performance of all four combinations: the two spectrograph configurations, quasi-Littrow (QL) and White Pupil (WP), with the two camera configurations, four lenses and five lenses. We demonstrate the strehl ratio in Fig. \ref{fig:Strehl_ratio}. This is a comparison of our combinations through the entire Y-band working wavelength. The quasi-Littrow configuration exceeds $>$ 80\% with either camera. However, it performs better with the 5 lenses combination. The White Pupil design shows a more symmetrical behaviour throughout the working wavelength range. It yields $>$ 90\% with both camera designs, with the 4-lens camera performing best. Thus, we choose the White Pupil configuration with the 4-lens camera as our final design.

%\newpage
\section{Analysis}
We investigate the robustness of our design to manufacturing and alignment defects (\S 3.1), thermal changes (\S 3.2).

\subsection{Tolerance Analysis}

The tolerance analysis focuses on the camera part of the spectrometer. This is a critical part of the system because the camera must be cooled down to 80K in a cryogenic Dewar. We reduced fabrication cost and delivery time by performing a testplate fit to all of the lenses. The testplate fit did not change the optical performance significantly, which is also presented in the previous section.

In the tolerance analysis, we forced ray aiming in ZEMAX for reliable results. For the alignment tolerances, the doublet was treated as a group. Additionally, we allowed the lenses to pivot about the center of curvature of the mating surface, to account for tolerances during cementing. The summary of the general tolerance variation is shown in Table  \ref{tab:tolerance}. 

\begin{figure}[h!]
\centering\includegraphics[scale=0.35]{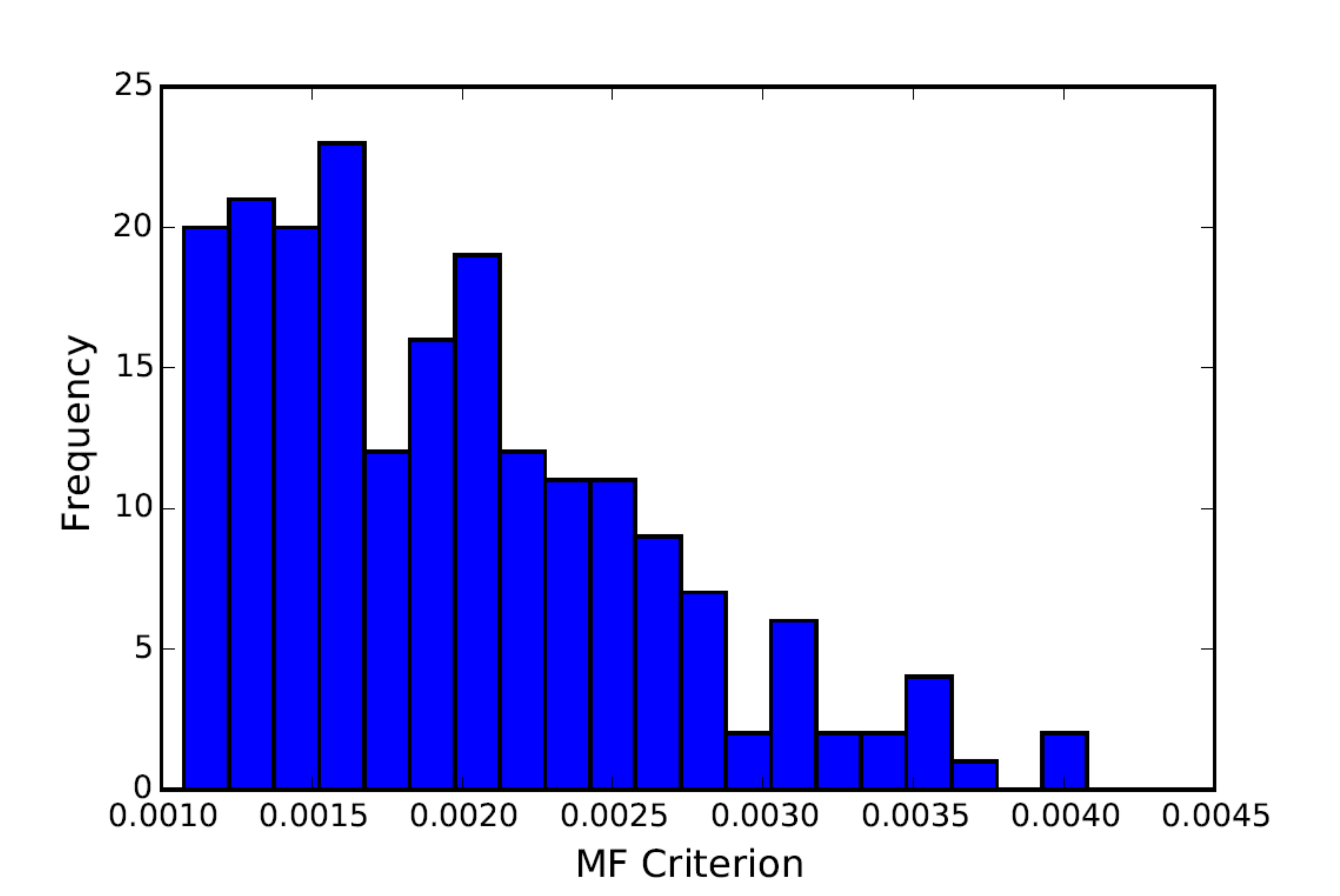}
\caption{Histogram of the merit value of 200 Monte Carlo draws of design within the tolerance criteria from Table. \ref{tab:tolerance}.}
\label{fig:Tol_analysis}
\end{figure}

We ran our analysis at a test wavelength of 1 $\mu$m.  We use detector tilt and focus as compensators, and optimize the design using Damped Least Squares. In a Monte Carlo procedure, ZEMAX generates lens tolerances randomly from within the assigned range which we list in Table. \ref{tab:tolerance}. The merit function measures the agreement between data and the fitting model for a particular choice of the parameters. By convention, the merit function is small when the agreement is good. The distribution of merit function values for 200 trials is shown in Fig \ref{fig:Tol_analysis}. As an example for a typical as-built system, we choose the design that lies at the mean of this distribution.

\begin{table}[ht!]
\caption {Camera Tolerance Criteria: parameters ranges to Monte Carlo generate 200 different designs.} \label{tab:tolerance} 
\begin{center}
\begin{tabular}{ c|c|c }
\hline
Surface & Property & Value \\
\hline
\multirow{1}{4em}{Doublet} &	Surface radius &	1 fringe  \\ 
& Center thickness &	$\pm$0.05 mm  \\ 
& Wedge &	$\pm$0.01 mm  \\ 
& Irregularity &	$\lambda$/8 mm  \\ 
& Tilt &	$\pm$0.2 mm  \\
& Decenter &	$\pm$0.2 mm  \\  
\hline
Group Spacing & Center thickness &	$\pm$0.05 mm  \\ 
\hline
\multirow{1}{4em}{Doublet} &	Surface radius &	1 fringe  \\
& Center thickness &	$\pm$0.05 mm  \\ 
& Wedge &	$\pm$0.01 mm  \\ 
& Irregularity &	$\lambda$/8 mm  \\ 
& Surface Tilt &	$\pm$0.2 mm  \\
& Surface Decenter &	$\pm$0.05 mm  \\ 
& Element Tilt &	$\pm$0.2 mm  \\ 
& Element Decenter &	$\pm$0.2 mm  \\ 
\hline
\end{tabular}
\end{center}
\end{table}

\subsection{Thermal Analysis}
The system environment of TARdYS' original design was set at a temperature of 20$^{\circ}$C and pressure 1 atm. As the last lens is cooled to cryogenic temperatures, we analysed the optical performance for the singlet L4 at 73K (-200~$^{\circ}$C).  
To obtain a reliable refractive index at cryogenic temperatures, we used the measurements of the Cryogenic, High-Accuracy Refraction Measuring System (CHARMS) \cite{2008arXiv0805.0091L}. They measured the absolute refractive index of fused silica from temperature ranging from 30 to 310 K at wavelength 0.4 to 2.6 $\mu$m with an absolute uncertainty of $\pm 1 \times 10^{-5}$. We contruct a new glass catalog with the Sellmeier 1 formula via the ZEMAX glass fitting tool.

From the glass fitting, we retrieve the new glass properties of Fused Silica and assign it for thermal analysis at -200 $^{\circ}$C. Initially the spot diagram had an airy disk size of 8.5 $\mu$m. The original setup shows a defocused spot diagram in the cold environment. The focusing problem can be solved by increasing the distance between the last Fused Silica lens and the detector mechanically. We optimized the thickness of the surface by using a merit function that considers all orders. The required detector tilt is 0.26 $^{\circ}$ in the cross dispersion direction. For an optimally focused setup, the optics are diffraction limited over the whole field.

%\newpage
\section{Prototype}
In this early phase of the project, we have built prototypes of the two different configurations according to our optical design. The main goal of this prototype is to test the optical components at ambient temperature. We exclude the image slicer from our experiment since its purpose is to boost up the spectral resolution of the spectrograph and the result has already been shown in our previous studies, including \cite{2017ExA....43..167T} and \cite{2018arXiv180407441V}.

\subsection{Optomechanical Design}
\begin{figure}[h!]
\centering\includegraphics[scale=0.4]{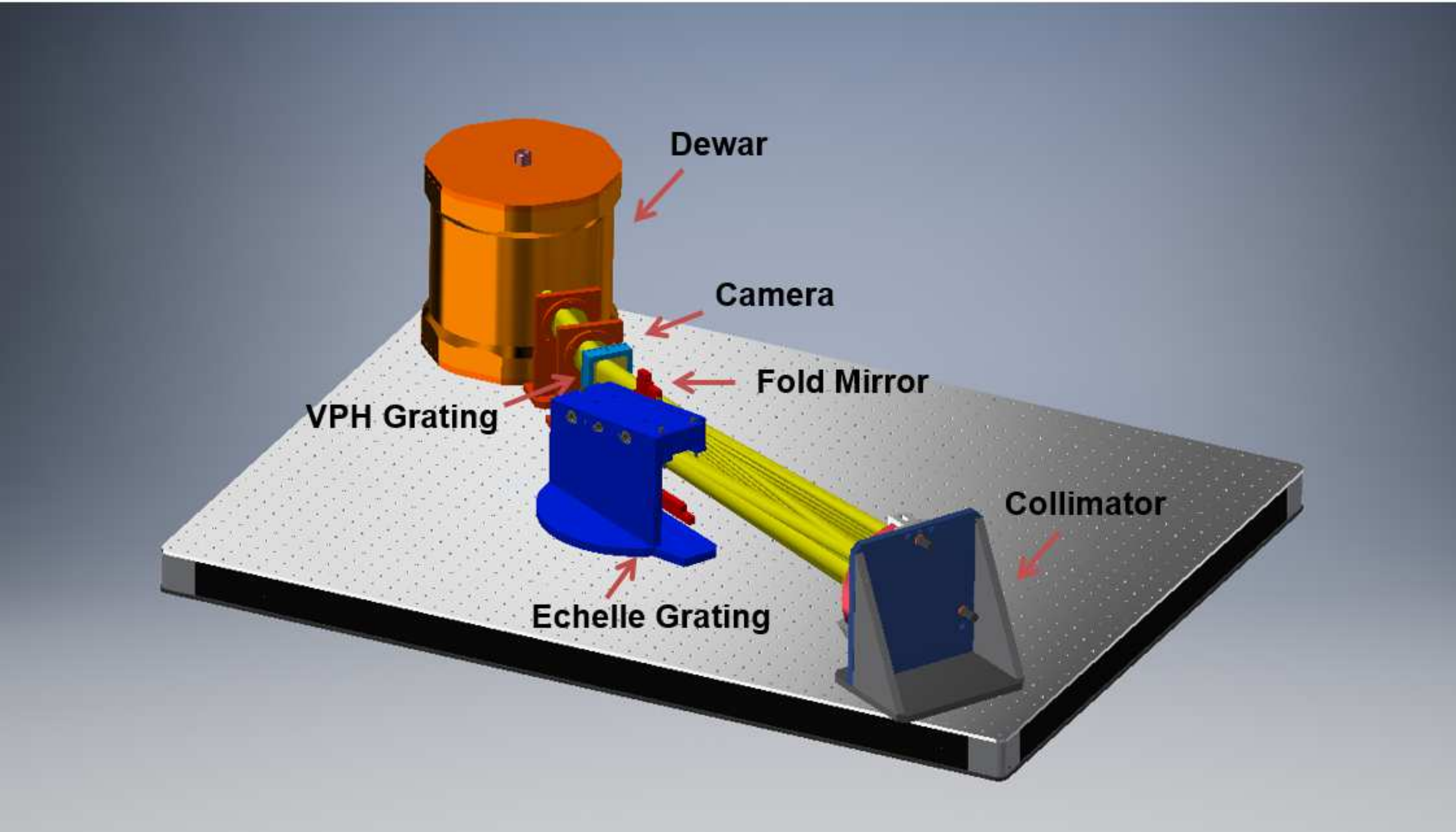}
\caption{Overview of the optomechanical design of the spectrograph in White Pupil Configuration in the real setup scenario. Here we include a bronze cryogenic Dewar that will cover the last lens of the system and the H1RG infrared detector.}
\label{fig:optomec}
\end{figure}

Compared to other spectrographs, TARdYS has a very compact design. As a result, the space constraints of TARdYS introduces some difficulties for designing optomechanical mounts. To do so, we first determine the available free space for every surface from the optical design footprint. We then determine the remaining space left for the mount. A general rule that we use for these mounts is the six points kinematic scheme. The overview of the TARdYS optomechanical design is displayed in Fig. \ref{fig:optomec}.

The collimating mirror a 156 mm diameter sized parabolic mirror with a focal length of 550 mm. It is made with ZERODUR and gold coated. The collimator mount housing structure is adapted from our previous spectrograph \cite{2014SPIE.9147E..89T}, and the back part of the collimator housing plate contains three micrometric screws that allow tip-tilt adjustments. The mirror's outer ring is adapted from the circular fold mirror mount of \cite{article}. 

The Echelle grating is placed facing down toward the optical table. We use a \textit{box} structure to hold the grating. We include the six points kinematic structure by applying hard points inside the box using end-hemispheres. A similar scheme is applied in \cite{unknown}. In our design, we cut down part of one wall, which is located close to input fiber, to avoid collision with the input optics. The housing structure allows a $\pm 2$ degrees tip-tilt of the grating box. For alignment, we introduce a hemisphere based plate (gray plate) purpose as shown in Fig. \ref{fig:optomec}.

The VPH grating works in transmission, thus its mount allows the light to pass through. The mount consists of 4 parts: mount frame, We designed a squared mount frame of the equal size to the grating, with three radial hard points and a spring plunger. It also has three axial hard points with spring forces pushing against the inside of the mount frame. The mount holder is designed so that the VPH grating has the same height as the optical path. The base plate is a circular plate that allows rotation perpendicular to the optical axis.

The fold mirror mount is designed to avoid vignetting in the crowded area for the White Pupil configuration. It applies a similar scheme as the VPH mount of rectangular shape optics. It also allows some small tip-tilt adjustments along the optical path direction. We use small rectangular pads opposite the three mount plungers.

The designs of the lens cells/frames are adjusted according to thickness of each lens. Each of them is attached to the mount frame which can be separated into two parts: frame and base. A spring frame is used to retain the lens inside the cell and to give small spring force against the cell. All three lenses of the camera part of the spectrograph have the same mount frame and spring frame.

\subsection{Alignment and construction}
\begin{figure}[h!]
\centering\includegraphics[scale=0.5]{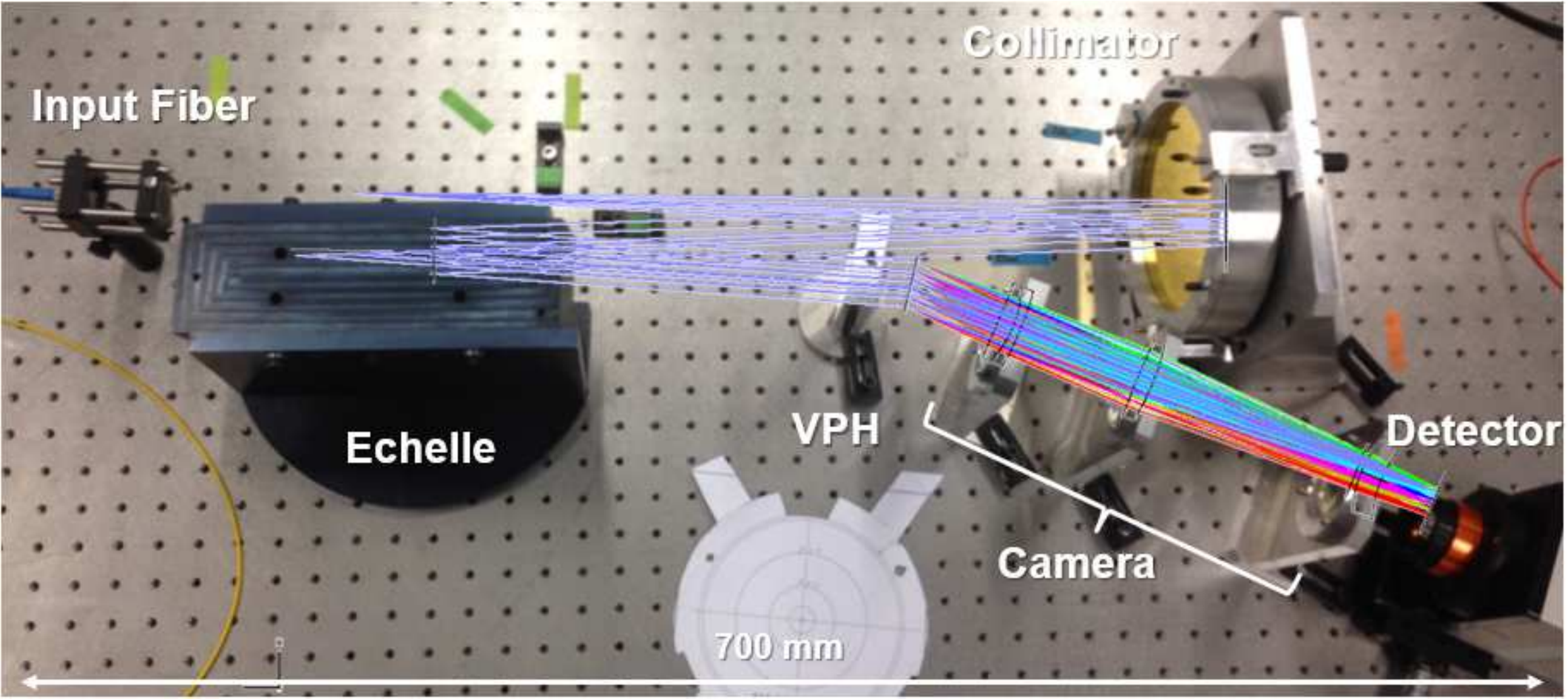}
\caption{The spectrograph prototype of the Littrow configuration overlayed with the ray tracing image, yielding the length of approximately 700 mm.}
\label{fig:prototypelittrow}
\end{figure}

In the quasi-littrow configuration, the light path starts from the object, goes to the collimator, Echelle grating, VPH grating and camera lenses, according to the optical design as shown in Fig. \ref{fig:prototypelittrow}. We also fabricate simple spacers between the camera lenses. In order to eliminate aberrations, we first use single mode fiber for the input and detect the images using a detector with a small pixel size such as SBIG detector. Later, we use the multi-mode fiber to obtain spectral images.

\begin{figure}[h!]
\centering\includegraphics[scale=0.5]{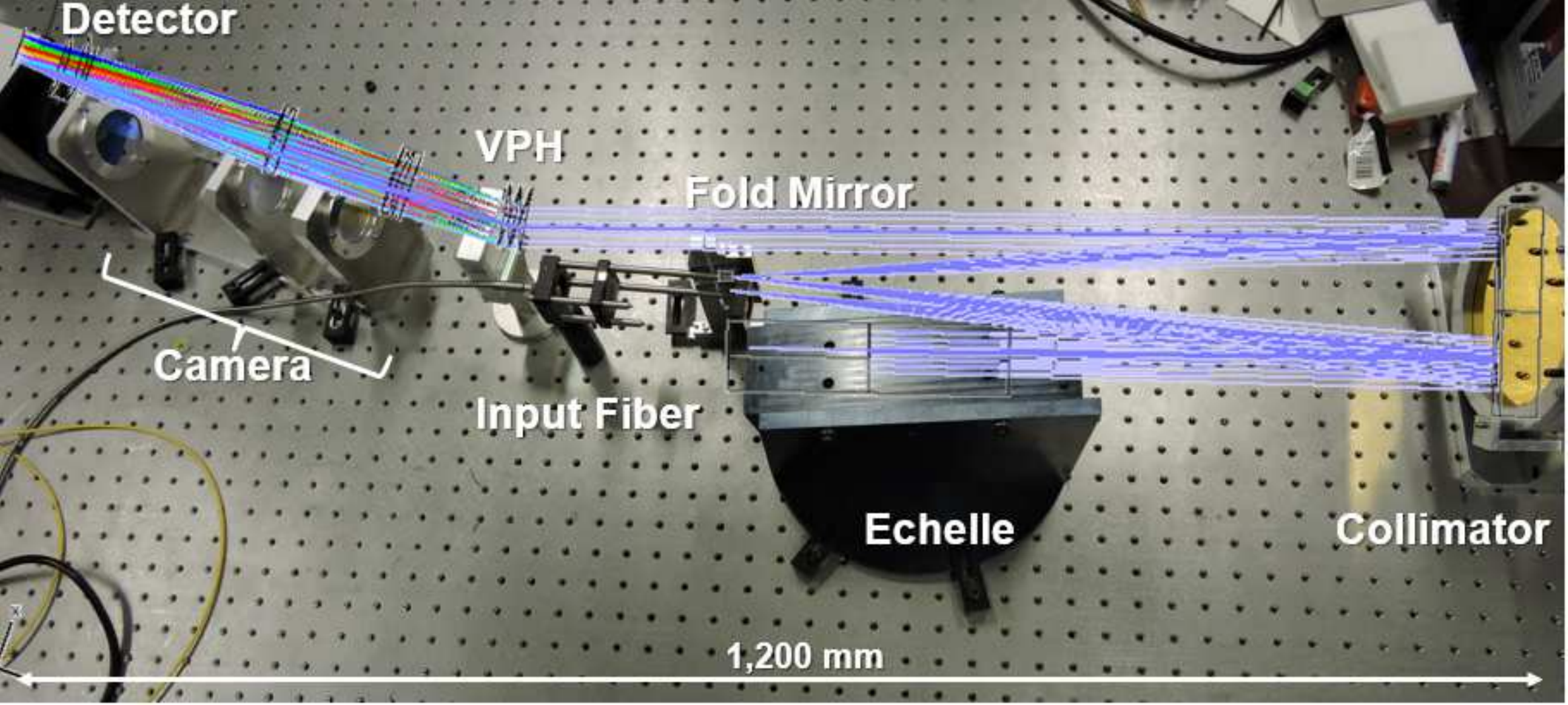}
\caption{The spectrograph prototype of the White Pupil configuration overlayed with the ray tracing image showing the double pass on the same collimating mirror. The length of the this configuration reaches approximately 1,200 mm after adjusting the optical design.}
\label{fig:prototypewhitepupil}
\end{figure}

In Fig.\ref{fig:prototypewhitepupil}, the Echelle, fold mirror, input optics and VPH grating are located in the limited space area from the original optical design of the White Pupil configuration. We adjust the optical design slightly to have a slightly larger angle of the beam reflected on the collimating mirror. We move the Echelle grating closer to the collimator and shift the VPH grating about 30 mm to the back. This results in a decrease in the strehl ratio (Fig. \ref{fig:Fig1B} bottom part) of about 3-5$\%$. The light path goes to the collimator, the Echelle, double passes on the collimator, goes to the fold mirror, and again hits the same collimator before going to the VPH cross disperser. Then the dispersed light goes through the camera lenses and detector. We overlay the ray tracing on the image of the real setup in Fig. \ref{fig:prototypewhitepupil}. One drawback of the WP prototype is the observable stray light on the image plane, for which we can possibly compensate by applying some baffles inside the spectrograph and anodizing the optomechanical mounts.

\subsection{Image Quality}

\begin{figure}[h!]
\centering
\includegraphics[height=4cm]{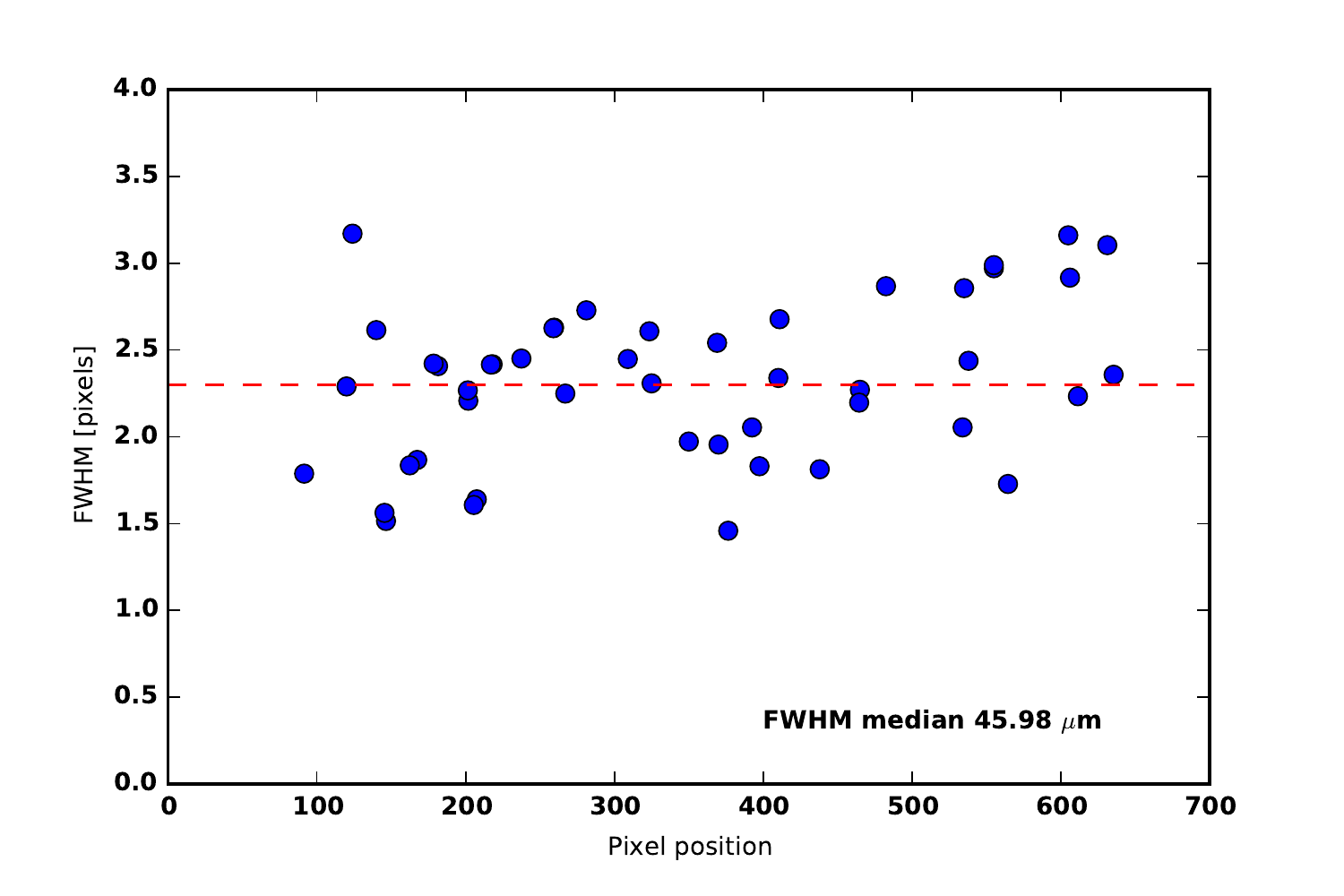}
\includegraphics[height=4cm]{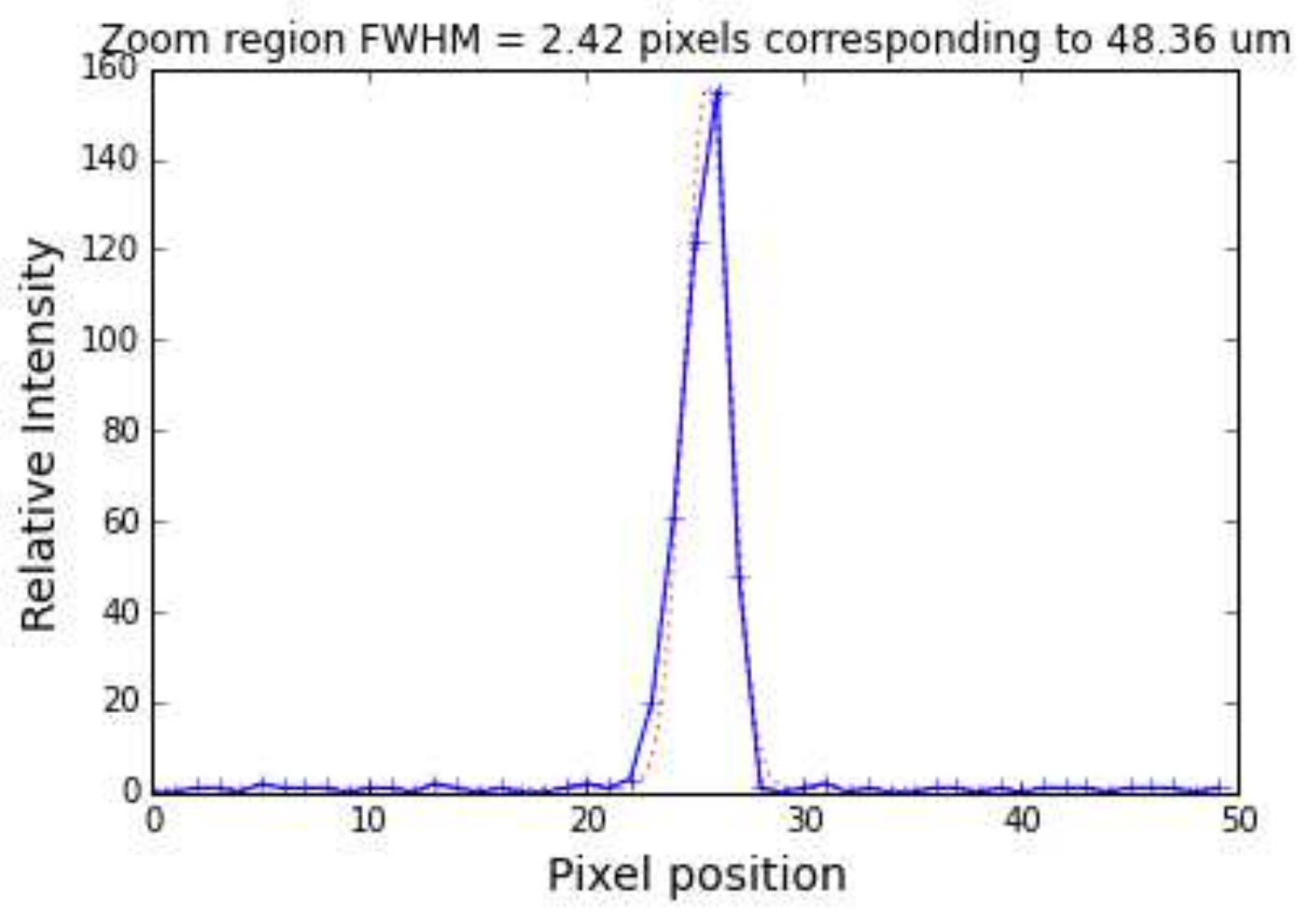}
\includegraphics[height=4cm]{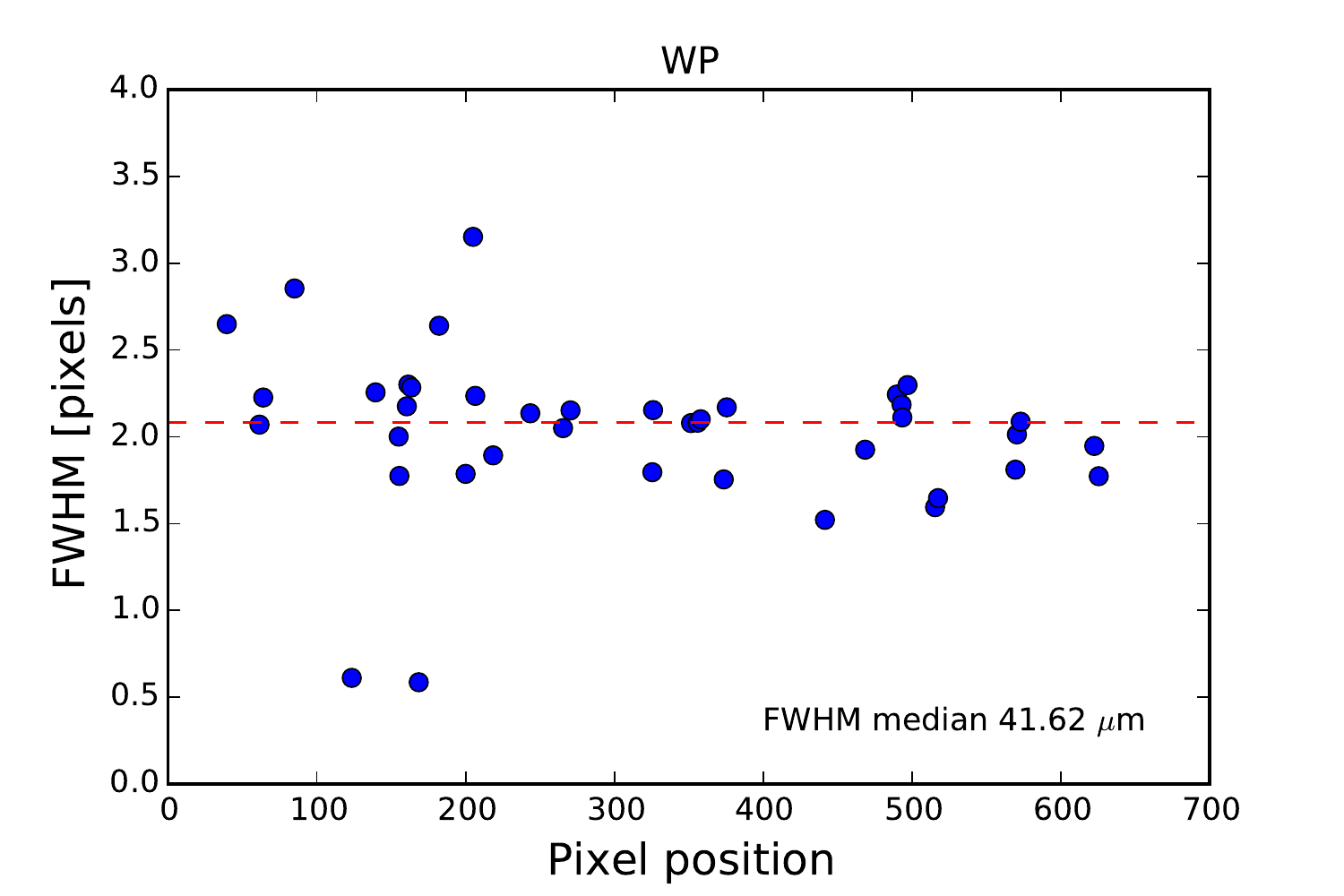}
\includegraphics[height=4cm]{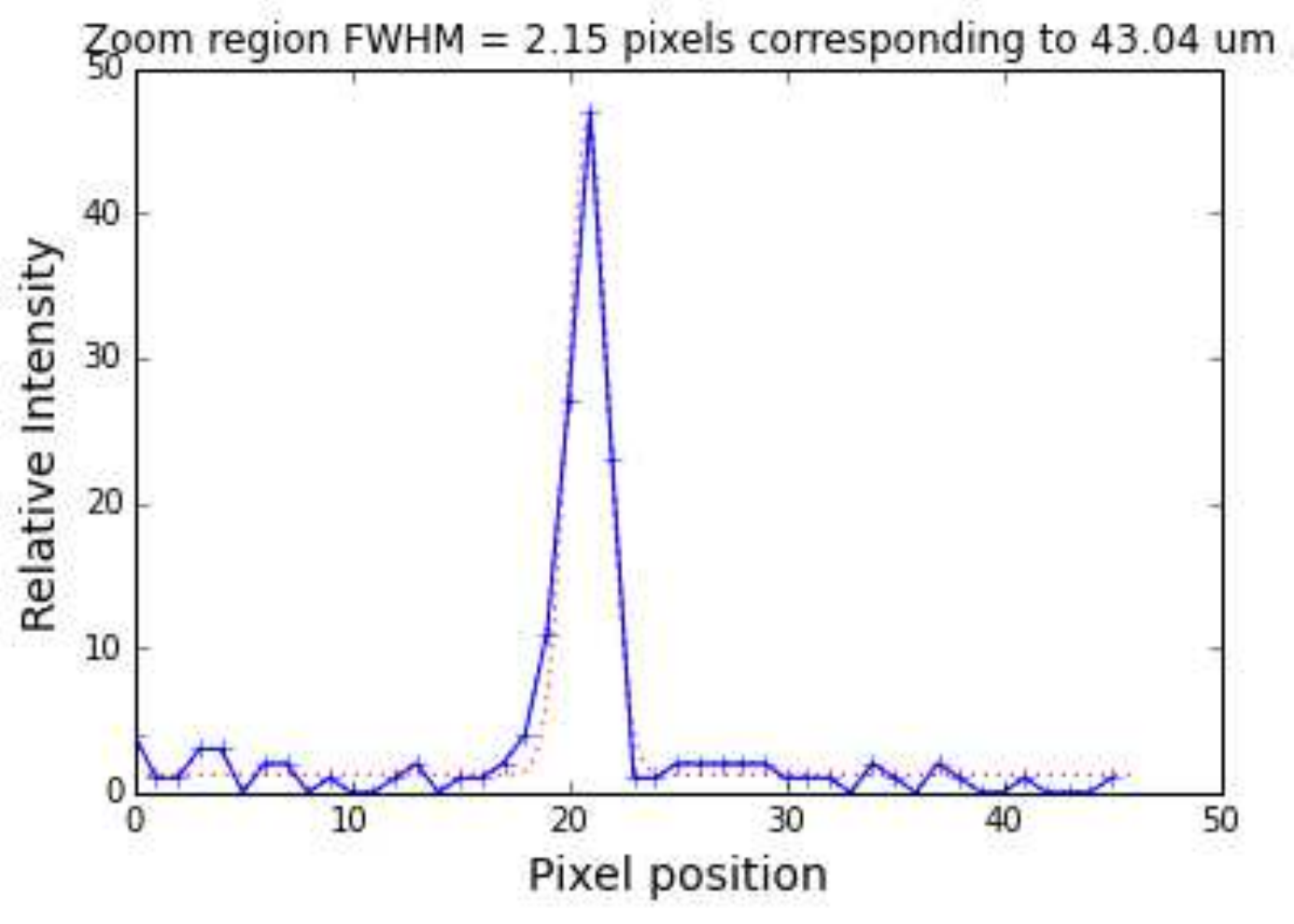}
\caption{Left: The FWHM analysis of one corner of the mosaic patch.  This analysis yields median FWHM of 2.4 pixels corresponding to approximately 48 $\mu$m from the QL configuration (top) and 2.15 pixels corresponding to 43 $\mu$m from WP configuration (bottom). Right: Typical PSF of calibration spot of the field showing the Gaussian fit (red dotted line) of the PSF which also fits the background level simultaneously.}
\label{fig:fwhm}
\end{figure}

We analyze the image quality from both prototype setups by taking spectral images with a ThAr calibration lamp using using a XEVA detector, which has a size of 512x640 pixels, and thus covers only a corner of the full IR detector field. In Fig. \ref{fig:fwhm}, we show FWHM of each spot in one such field. Across all spots, the median FWHM is 2.4 pixels which corresponds to  48 $\mu$m. This can be translated to the spectral resolution R = 35,617 at, for example, the 160th order. In the WP configuration, the median FWHM yields 2.15 pixels, corresponding to 43 $\mu$m. This improves the image quality by approximately 10\%.

\newpage
\subsection{Stability}

For this prototype experiment, we control the temperature with a very simple method. Our goal is to keep the room, detector and optical bench stable within 1 degree. We control the room temperature with an air-conditioner (AC) set to 22 $^\circ$C in cold mode during the summer and heat mode during the winter. Without any enclosure, all the spectrograph parts are exposed to the room temperature. One problem of controlling with an AC is the fluctuation due to its cycle, in our case $ \pm 1 ^\circ$C  approximately every 20 minutes. This fluctuation has a direct effect on the XEVA infrared detector, which we use for this prototype, because XEVA applies electronic cooling (TE3) to cool down using an offset temperature from the ambient environment. Another problem of using this detector is the heat generated from the detector, as it heats up the optical bench over time. We test different conditions as described in appendix. 

%\subsection{Shift/Drift}
Together with the temperature control experiment, we also took data from one corner of the full field (see appendix C) in order to measure the wavelength stability. In this experiment the room temperature is controlled, while the pressure is not. We use a semi-open enclosure and monitor the wavelength shift over 4 hours. Data sets are obtained every 30 minutes.

\begin{figure}[ht!]
\centering
\includegraphics[width=0.33\textwidth]{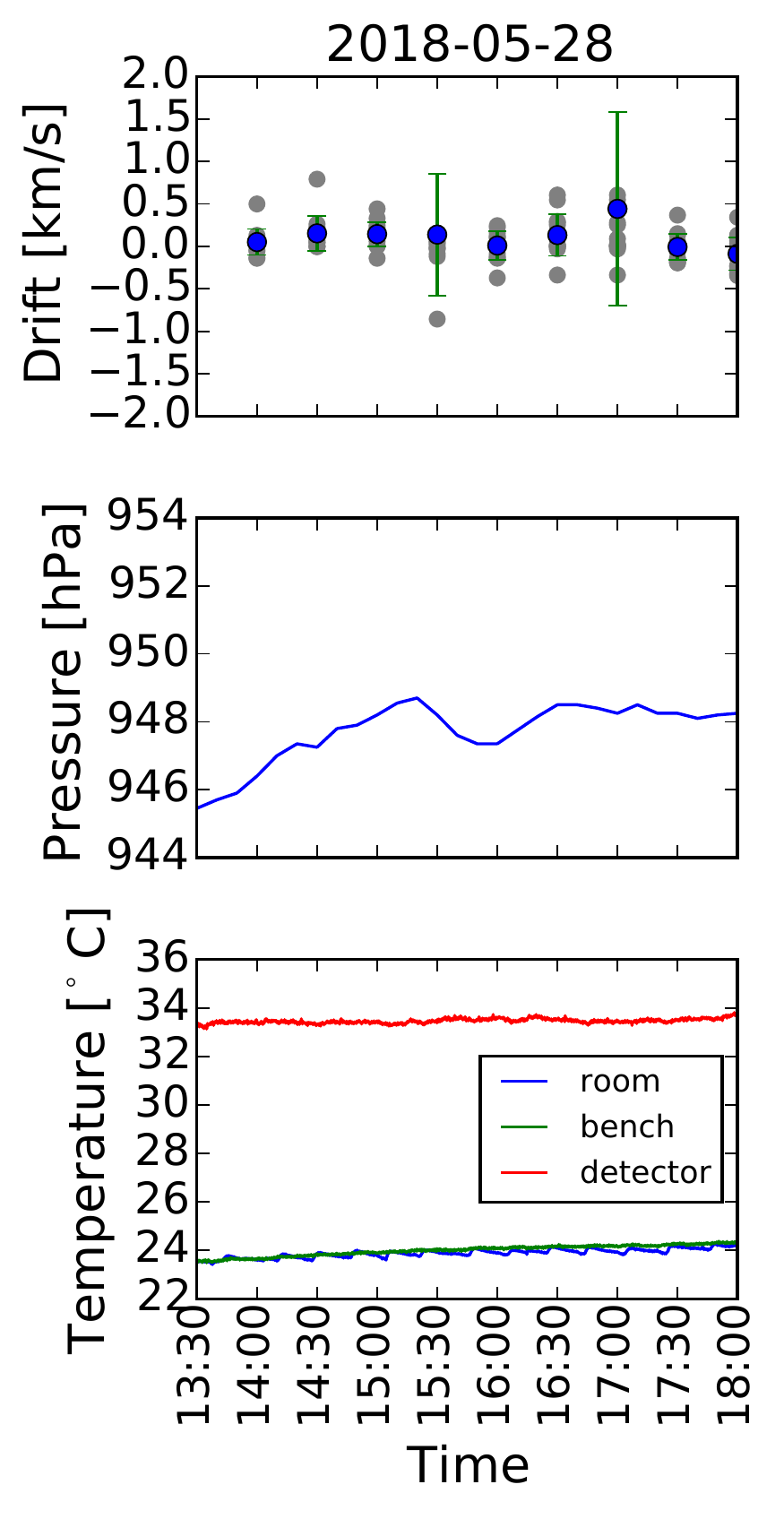}\includegraphics[width=0.33\textwidth]{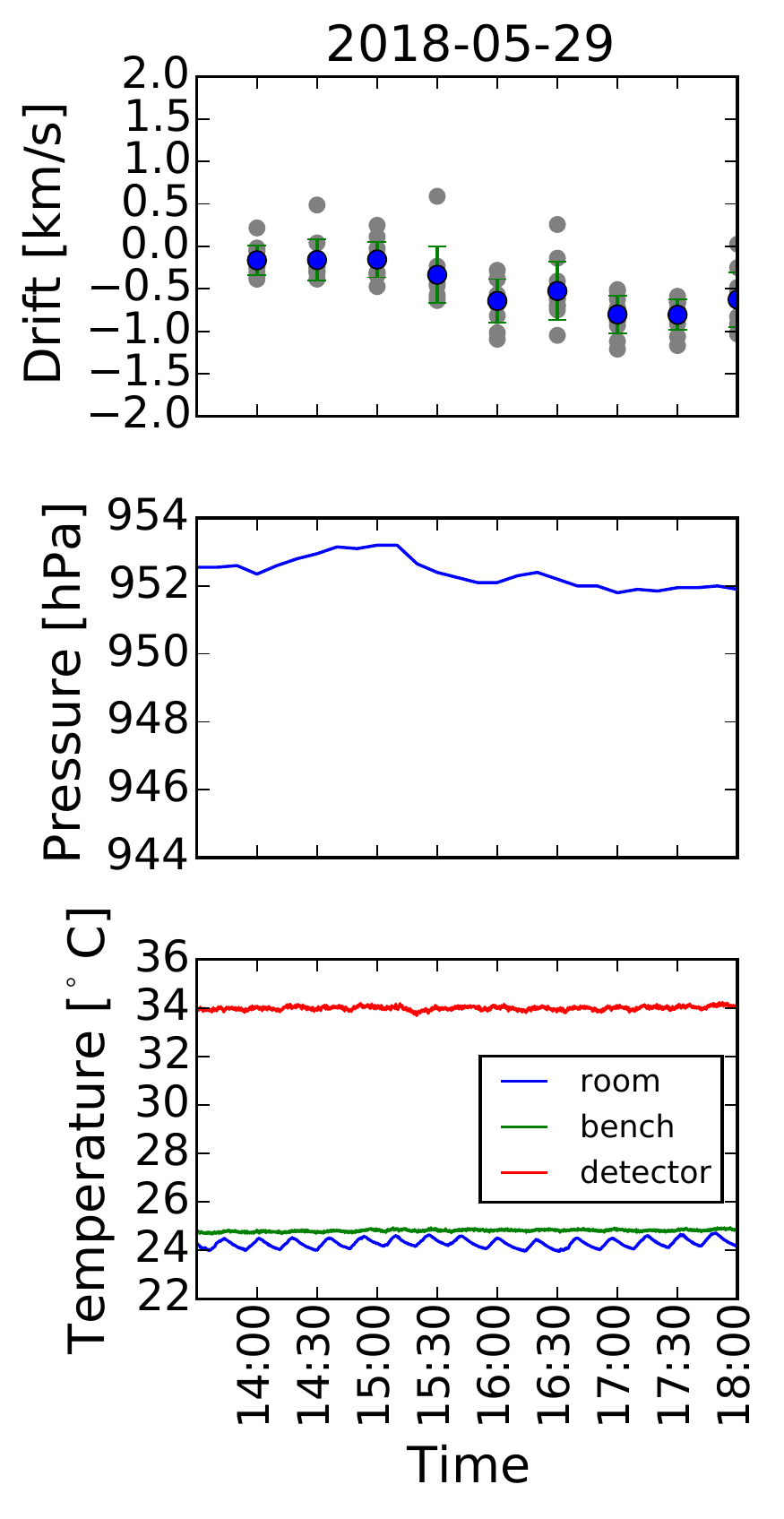}\includegraphics[width=0.33\textwidth]{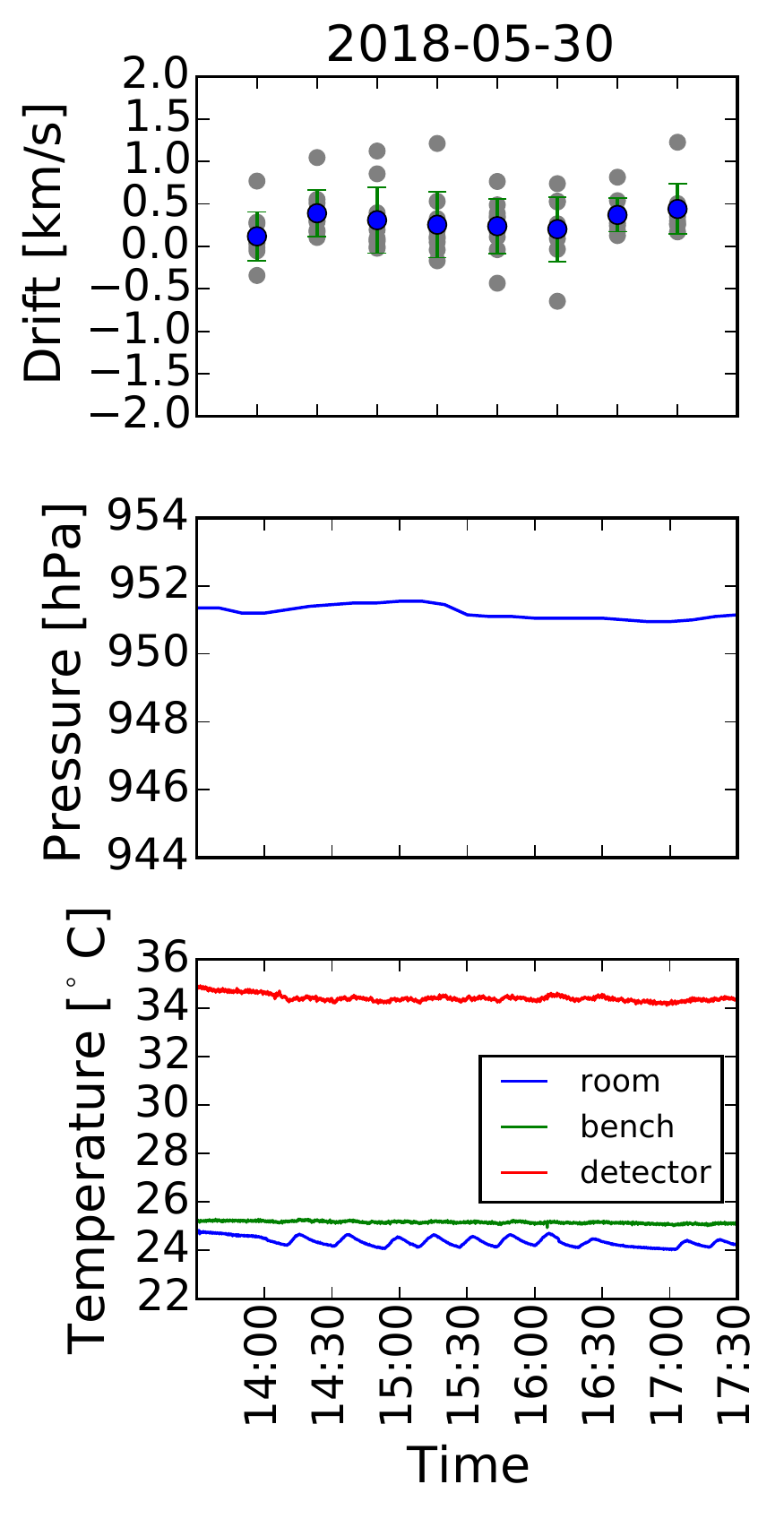}%
\caption{The Radial Velocity shift of the ThAr lamp in the AC control environment with enclosure. In the top panel, the gray dots represent sampled data points, and the blue dots represent average value representatives of each data set with green error bar. The middle panel is the measured pressure from the nearby pressure station. The bottom panel is temperature behavior during the measurement.}
\label{fig:rvshift3}
\end{figure}

We found that stabilizing temperature of the detector and the optical bench helps decrease the amplitude of the drift. At a $ \pm 0.1 ^\circ$C control level, the drift varies the amplitude within 1 pixel of 20 $\mu$m corresponding  to $\approx$ 2 km/s peak to peak in the top panel of Fig. \ref{fig:rvshift3}. Within this range, one can observe that the drift is mostly affected by the environmental pressure (middle panel). This same effect was also observed in our previous study \cite{2018arXiv180407441V}.

%\newpage
\section{Conclusion}
We present the optical design and prototype of TARdYS, a fiber-fed high resolution near-infrared spectrograph for RV exoplanet search. Our novel design uses for the first time an Echelle R6 and an image-slicer.

We explore and compare the designs of two popular spectrograph setups: quasi-Littrow and White Pupil and two customized cameras. The simple quasi-Littrow configuration shows that it can yield a relatively high strehl ratio with a proper camera design. The Huygen PSF analysis shows a slightly tilted image on the detector plane due to the setup. However, to detect very small shifts on spectral lines of exoplanets, we require maximum throughput, which we can obtain from the White Pupil design with the same budget by double passing the light on the same collimating mirror. The Echelle R6, compared with other Echelle types, gives the smallest angle to the instrument by far. This allows us to shrink the volume of the optomechanical mounts and bench, resulting in a compact design. Our design packs 42 spectral orders covering Y band (0.843-1.117 $\mu$m) onto a 1024x1024 pixels infrared detector, which is the most costly component of the instrument. With an image slicer, it potentially yields spectral resolution of R = 60,000.

The final design of TARdYS, based on the White Pupil configuration, was presented. The tolerance and thermal analysis of the spectrograph camera predicts diffraction limited performance assuming realistic manufacturing and alignment tolerances. The spectrograph prototype was built in the lab Center of Astro-Engineering UC-AIUC in Chile. We have retrieved preliminary results from our prototype setup using a commercially available infrared detector XEVA. We obtained mosaic a full field image of the Thorium-Argon spectra and IR continuum source using a pattern recognition technique.  The result agrees well with the simulation, given the median image quality of 2.4 pixels FWHM corresponding to spectral resolution of approximately 35,000 without an image slicer. The image quality improve by 8-10\% in the WP configuration. We also applied a simple temperature stability control to the prototype using an air conditioner and a semi-open enclosure to circulate the heat within the system. This results in a change of $\pm$ 0.1 K during over several hours. We found that stabilizing the temperature of the detector and optical bench temperature helps to decrease the amplitude of the velocity drift. Within the range of 2 km/s, the trend of the drift is most affected by the environmental pressure.

The next phase of the project will include the integration of the detector, Dewar and a better-structured stability control. We plan to use a Uranium lamp as a calibrator instead of Thorium-Argon combining with Fabry-Perot Etalon. TARdYS will open up opportunities for high-resolution Y-band infrared spectroscopy including studies of M-dwarfs and searches for the exoplanets.

%
%

%\newpage
\begin{acknowledgements}
S.R. acknowledges support from Direccion de Doctorado de la Vicerrectoria de Investigacion,(VRI), Pontificia Universidad Catolica de Chile. She thanks CONICYT fondo QUIMAL that will eventually fund the 1K infrared detector and Dewar and make this project become a real NIR exoplanet hunter in the near future. She thanks Andrei A. Tokovinin for suggestions for the project. S.R. thanks Johannes Buchner for his advice on the technical programming issues. We thank Johannes Buchner and Abigail Stein for proof-reading.
\end{acknowledgements}

% BibTeX users please use one of
%\bibliographystyle{spbasic}      % basic style, author-year citations
%\bibliographystyle{spmpsci}      % mathematics and physical sciences
\bibliographystyle{spphys}       % APS-like style for physics
\bibliography{ref}   % name your BibTeX data base

\begin{thebibliography}{10}
\providecommand{\url}[1]{{#1}}
\providecommand{\urlprefix}{URL }
\expandafter\ifx\csname urlstyle\endcsname\relax
  \providecommand{\doi}[1]{DOI \discretionary{}{}{}#1}\else
  \providecommand{\doi}{DOI \discretionary{}{}{}\begingroup
  \urlstyle{rm}\Url}\fi

\bibitem{1995Natur.378..355M}
M.~{Mayor}, D.~{Queloz}, {A Jupiter-mass companion to a solar-type star}, \nat
  \textbf{378}, 355 (1995).
\newblock \doi{10.1038/378355a0}

\bibitem{2011exha.book.....P}
M.~{Perryman}, \emph{{The Exoplanet Handbook}} (2011)

\bibitem{2016ApJS..226....7C}
I.J.M. {Crossfield}, D.R. {Ciardi}, E.A. {Petigura}, E.~{Sinukoff}, J.E.
  {Schlieder}, A.W. {Howard}, C.A. {Beichman}, H.~{Isaacson}, C.D. {Dressing},
  J.L. {Christiansen}, B.J. {Fulton}, S.~{L{\'e}pine}, L.~{Weiss}, L.~{Hirsch},
  J.~{Livingston}, C.~{Baranec}, N.M. {Law}, R.~{Riddle}, C.~{Ziegler}, S.B.
  {Howell}, E.~{Horch}, M.~{Everett}, J.~{Teske}, A.O. {Martinez},
  C.~{Obermeier}, B.~{Benneke}, N.~{Scott}, N.~{Deacon}, K.M. {Aller}, B.M.S.
  {Hansen}, L.~{Mancini}, S.~{Ciceri}, R.~{Brahm}, A.~{Jord{\'a}n}, H.A.
  {Knutson}, T.~{Henning}, M.~{Bonnefoy}, M.C. {Liu}, J.R. {Crepp},
  J.~{Lothringer}, P.~{Hinz}, V.~{Bailey}, A.~{Skemer}, D.~{Defrere}, {197
  Candidates and 104 Validated Planets in K2's First Five Fields}, \apjs
  \textbf{226}, 7 (2016).
\newblock \doi{10.3847/0067-0049/226/1/7}

\bibitem{2016ApJS..222...14V}
A.~{Vanderburg}, D.W. {Latham}, L.A. {Buchhave}, A.~{Bieryla}, P.~{Berlind},
  M.L. {Calkins}, G.A. {Esquerdo}, S.~{Welsh}, J.A. {Johnson}, {Planetary
  Candidates from the First Year of the K2 Mission}, \apjs \textbf{222}, 14
  (2016).
\newblock \doi{10.3847/0067-0049/222/1/14}

\bibitem{2007AsBio...7...30T}
J.C. {Tarter}, P.R. {Backus}, R.L. {Mancinelli}, J.M. {Aurnou}, D.E. {Backman},
  G.S. {Basri}, A.P. {Boss}, A.~{Clarke}, D.~{Deming}, L.R. {Doyle}, E.D.
  {Feigelson}, F.~{Freund}, D.H. {Grinspoon}, R.M. {Haberle}, S.A. {Hauck}, II,
  M.J. {Heath}, T.J. {Henry}, J.L. {Hollingsworth}, M.M. {Joshi}, S.~{Kilston},
  M.C. {Liu}, E.~{Meikle}, I.N. {Reid}, L.J. {Rothschild}, J.~{Scalo},
  A.~{Segura}, C.M. {Tang}, J.M. {Tiedje}, M.C. {Turnbull}, L.M. {Walkowicz},
  A.L. {Weber}, R.E. {Young}, {A Reappraisal of The Habitability of Planets
  around M Dwarf Stars}, Astrobiology \textbf{7}, 30 (2007).
\newblock \doi{10.1089/ast.2006.0124}

\bibitem{2010ApJ...710..432R}
A.~{Reiners}, J.L. {Bean}, K.F. {Huber}, S.~{Dreizler}, A.~{Seifahrt},
  S.~{Czesla}, {Detecting Planets Around Very Low Mass Stars with the Radial
  Velocity Method}, \apj \textbf{710}, 432 (2010).
\newblock \doi{10.1088/0004-637X/710/1/432}

\bibitem{2014Natur.513..358P}
F.~{Pepe}, D.~{Ehrenreich}, M.R. {Meyer}, {Instrumentation for the detection
  and characterization of exoplanets}, \nat \textbf{513}, 358 (2014).
\newblock \doi{10.1038/nature13784}

\bibitem{Moorwood2005}
A.~Moorwood, \emph{CRIRES: Context and Status} (Springer Berlin Heidelberg,
  Berlin, Heidelberg, 2005), pp. 15--24.
\newblock \doi{10.1007/10995082\_2}.
\newblock \urlprefix\url{https://doi.org/10.1007/10995082\_2}

\bibitem{2006SPIE.6269E..19O}
E.~{Oliva}, L.~{Origlia}, C.~{Baffa}, C.~{Biliotti}, P.~{Bruno}, F.~{D'Amato},
  C.~{Del Vecchio}, G.~{Falcini}, S.~{Gennari}, F.~{Ghinassi}, E.~{Giani},
  M.~{Gonzalez}, F.~{Leone}, M.~{Lolli}, M.~{Lodi}, R.~{Maiolino},
  F.~{Mannucci}, G.~{Marcucci}, I.~{Mochi}, P.~{Montegriffo}, E.~{Rossetti},
  S.~{Scuderi}, M.~{Sozzi}, in \emph{Society of Photo-Optical Instrumentation
  Engineers (SPIE) Conference Series}, \emph{\procspie}, vol. 6269 (2006),
  \emph{\procspie}, vol. 6269, p. 626919.
\newblock \doi{10.1117/12.670006}

\bibitem{2010SPIE.7735E..1MY}
I.S. {Yuk}, D.T. {Jaffe}, S.~{Barnes}, M.Y. {Chun}, C.~{Park}, S.~{Lee},
  H.~{Lee}, W.~{Wang}, K.J. {Park}, S.~{Pak}, J.~{Strubhar}, C.~{Deen},
  H.~{Oh}, H.~{Seo}, T.S. {Pyo}, W.K. {Park}, J.~{Lacy}, J.~{Goertz},
  J.~{Rand}, M.~{Gully-Santiago}, in \emph{Ground-based and Airborne
  Instrumentation for Astronomy III}, \emph{\procspie}, vol. 7735 (2010),
  \emph{\procspie}, vol. 7735, p. 77351M.
\newblock \doi{10.1117/12.856864}

\bibitem{2014SPIE.9147E..1GM}
S.~{Mahadevan}, L.W. {Ramsey}, R.~{Terrien}, S.~{Halverson}, A.~{Roy},
  F.~{Hearty}, E.~{Levi}, G.K. {Stefansson}, P.~{Robertson}, C.~{Bender},
  C.~{Schwab}, M.~{Nelson}, in \emph{Ground-based and Airborne Instrumentation
  for Astronomy V}, \emph{\procspie}, vol. 9147 (2014), \emph{\procspie}, vol.
  9147, p. 91471G.
\newblock \doi{10.1117/12.2056417}

\bibitem{2012SPIE.8446E..1TT}
M.~{Tamura}, H.~{Suto}, J.~{Nishikawa}, T.~{Kotani}, B.~{Sato}, W.~{Aoki},
  T.~{Usuda}, T.~{Kurokawa}, K.~{Kashiwagi}, S.~{Nishiyama}, Y.~{Ikeda},
  D.~{Hall}, K.~{Hodapp}, J.~{Hashimoto}, J.~{Morino}, S.~{Inoue}, Y.~{Mizuno},
  Y.~{Washizaki}, Y.~{Tanaka}, S.~{Suzuki}, J.~{Kwon}, T.~{Suenaga}, D.~{Oh},
  N.~{Narita}, E.~{Kokubo}, Y.~{Hayano}, H.~{Izumiura}, E.~{Kambe}, T.~{Kudo},
  N.~{Kusakabe}, M.~{Ikoma}, Y.~{Hori}, M.~{Omiya}, H.~{Genda}, A.~{Fukui},
  Y.~{Fujii}, O.~{Guyon}, H.~{Harakawa}, M.~{Hayashi}, M.~{Hidai}, T.~{Hirano},
  M.~{Kuzuhara}, M.~{Machida}, T.~{Matsuo}, T.~{Nagata}, H.~{Ohnuki},
  M.~{Ogihara}, S.~{Oshino}, R.~{Suzuki}, H.~{Takami}, N.~{Takato},
  Y.~{Takahashi}, C.~{Tachinami}, H.~{Terada}, in \emph{Ground-based and
  Airborne Instrumentation for Astronomy IV}, \emph{\procspie}, vol. 8446
  (2012), \emph{\procspie}, vol. 8446, p. 84461T.
\newblock \doi{10.1117/12.925885}

\bibitem{2014SPIE.9147E..1FQ}
A.~{Quirrenbach}, P.J. {Amado}, J.A. {Caballero}, R.~{Mundt}, A.~{Reiners},
  I.~{Ribas}, W.~{Seifert}, M.~{Abril}, J.~{Aceituno}, F.J. {Alonso-Floriano},
  M.~{Ammler-von Eiff}, R.~{Antona Jim{\'e}nez}, H.~{Anwand-Heerwart},
  M.~{Azzaro}, F.~{Bauer}, D.~{Barrado}, S.~{Becerril}, V.J.S. {B{\'e}jar},
  D.~{Ben{\'{\i}}tez}, Z.M. {Berdi{\~n}as}, M.C. {C{\'a}rdenas}, E.~{Casal},
  A.~{Claret}, J.~{Colom{\'e}}, M.~{Cort{\'e}s-Contreras}, S.~{Czesla},
  M.~{Doellinger}, S.~{Dreizler}, C.~{Feiz}, M.~{Fern{\'a}ndez},
  D.~{Galad{\'{\i}}}, M.C. {G{\'a}lvez-Ortiz}, A.~{Garc{\'{\i}}a-Piquer}, M.L.
  {Garc{\'{\i}}a-Vargas}, R.~{Garrido}, L.~{Gesa}, V.~{G{\'o}mez Galera},
  E.~{Gonz{\'a}lez {\'A}lvarez}, J.I. {Gonz{\'a}lez Hern{\'a}ndez},
  U.~{Gr{\"o}zinger}, J.~{Gu{\`a}rdia}, E.W. {Guenther}, E.~{de Guindos},
  J.~{Guti{\'e}rrez-Soto}, H.J. {Hagen}, A.P. {Hatzes}, P.H. {Hauschildt},
  J.~{Helmling}, T.~{Henning}, D.~{Hermann}, L.~{Hern{\'a}ndez Casta{\~n}o},
  E.~{Herrero}, D.~{Hidalgo}, G.~{Holgado}, A.~{Huber}, K.F. {Huber},
  S.~{Jeffers}, V.~{Joergens}, E.~{de Juan}, M.~{Kehr}, R.~{Klein},
  M.~{K{\"u}rster}, A.~{Lamert}, S.~{Lalitha}, W.~{Laun}, U.~{Lemke},
  R.~{Lenzen}, M.~{L{\'o}pez del Fresno}, B.~{L{\'o}pez Mart{\'{\i}}},
  J.~{L{\'o}pez-Santiago}, U.~{Mall}, H.~{Mandel}, E.L. {Mart{\'{\i}}n},
  S.~{Mart{\'{\i}}n-Ruiz}, H.~{Mart{\'{\i}}nez-Rodr{\'{\i}}guez}, C.J.
  {Marvin}, R.J. {Mathar}, E.~{Mirabet}, D.~{Montes}, R.~{Morales Mu{\~n}oz},
  A.~{Moya}, V.~{Naranjo}, A.~{Ofir}, R.~{Oreiro}, E.~{Pall{\'e}},
  J.~{Panduro}, V.M. {Passegger}, A.~{P{\'e}rez-Calpena}, D.~{P{\'e}rez
  Medialdea}, M.~{Perger}, M.~{Pluto}, A.~{Ram{\'o}n}, R.~{Rebolo},
  P.~{Redondo}, S.~{Reffert}, S.~{Reinhardt}, P.~{Rhode}, H.W. {Rix},
  F.~{Rodler}, E.~{Rodr{\'{\i}}guez}, C.~{Rodr{\'{\i}}guez-L{\'o}pez},
  E.~{Rodr{\'{\i}}guez-P{\'e}rez}, R.R. {Rohloff}, A.~{Rosich},
  E.~{S{\'a}nchez-Blanco}, M.A. {S{\'a}nchez Carrasco}, J.~{Sanz-Forcada}, L.F.
  {Sarmiento}, S.~{Sch{\"a}fer}, J.~{Schiller}, C.~{Schmidt}, J.H.M.M.
  {Schmitt}, E.~{Solano}, O.~{Stahl}, C.~{Storz}, J.~{St{\"u}rmer}, J.C.
  {Su{\'a}rez}, R.G. {Ulbrich}, G.~{Veredas}, K.~{Wagner}, J.~{Winkler}, M.R.
  {Zapatero Osorio}, M.~{Zechmeister}, F.J. {Abell{\'a}n de Paco},
  G.~{Anglada-Escud{\'e}}, C.~{del Burgo}, A.~{Klutsch}, J.L. {Lizon},
  M.~{L{\'o}pez-Morales}, J.C. {Morales}, M.A.C. {Perryman}, S.M. {Tulloch},
  W.~{Xu}, in \emph{Ground-based and Airborne Instrumentation for Astronomy V},
  \emph{\procspie}, vol. 9147 (2014), \emph{\procspie}, vol. 9147, p. 91471F.
\newblock \doi{10.1117/12.2056453}

\bibitem{2014SPIE.9147E..15A}
{\'E}.~{Artigau}, D.~{Kouach}, J.F. {Donati}, R.~{Doyon}, X.~{Delfosse},
  S.~{Baratchart}, M.~{Lacombe}, C.~{Moutou}, P.~{Rabou}, L.P. {Par{\`e}s},
  Y.~{Micheau}, S.~{Thibault}, V.A. {Reshetov}, B.~{Dubois}, O.~{Hernandez},
  P.~{Vall{\'e}e}, S.Y. {Wang}, F.~{Dolon}, F.A. {Pepe}, F.~{Bouchy},
  N.~{Striebig}, F.~{H{\'e}nault}, D.~{Loop}, L.~{Saddlemyer}, G.~{Barrick},
  T.~{Vermeulen}, M.~{Dupieux}, G.~{H{\'e}brard}, I.~{Boisse}, E.~{Martioli},
  S.H.P. {Alencar}, J.D. {do Nascimento}, P.~{Figueira}, in \emph{Ground-based
  and Airborne Instrumentation for Astronomy V}, \emph{\procspie}, vol. 9147
  (2014), \emph{\procspie}, vol. 9147, p. 914715.
\newblock \doi{10.1117/12.2055663}

\bibitem{2016SPIE.9908E..5ZI}
Y.~{Ikeda}, N.~{Kobayashi}, S.~{Kondo}, S.~{Otsubo}, S.~{Hamano},
  H.~{Sameshima}, T.~{Yoshikawa}, K.~{Fukue}, K.~{Nakanishi}, T.~{Kawanishi},
  T.~{Nakaoka}, M.~{Kinoshita}, A.~{Kitano}, A.~{Asano}, K.~{Takenaka},
  A.~{Watase}, H.~{Mito}, C.~{Yasui}, A.~{Minami}, N.~{Izumu}, R.~{Yamamoto},
  M.~{Mizumoto}, T.~{Arasaki}, A.~{Arai}, N.~{Matsunaga}, H.~{Kawakita}, in
  \emph{Ground-based and Airborne Instrumentation for Astronomy VI},
  \emph{\procspie}, vol. 9908 (2016), \emph{\procspie}, vol. 9908, p. 99085Z.
\newblock \doi{10.1117/12.2230886}

\bibitem{doi:10.1117/12.2231391}
Y.~{Yoshii}, M.~{Doi}, K.~{Kohno}, T.~{Miyata}, K.~{Motohara}, K.~{Kawara},
  M.~{Tanaka}, T.~{Minezaki}, S.~{Sako}, T.~{Morokuma}, Y.~{Tamura},
  T.~{Tanabe}, H.~{Takahashi}, M.~{Konishi}, T.~{Kamizuka}, N.~{Kato},
  T.~{Aoki}, T.~{Soyano}, K.~{Tarusawa}, T.~{Handa}, S.~{Koshida},
  L.~{Bronfman}, M.T. {Ruiz}, M.~{Hamuy}, G.~{Garay}.
\newblock The university of tokyo atacama observatory 6.5m telescope: project
  overview and current status (2016).
\newblock \doi{10.1117/12.2231391}.
\newblock \urlprefix\url{https://doi.org/10.1117/12.2231391}

\bibitem{1938ApJ....88..113B}
I.S. {Bowen}, {The Image-Slicer a Device for Reducing Loss of Light at Slit of
  Stellar Spectrograph.}, \apj \textbf{88}, 113 (1938).
\newblock \doi{10.1086/143964}

\bibitem{doi:10.1117/12.856709}
C.~{Schwab}, J.F.P. {Spronck}, A.~{Tokovinin}, D.A. {Fischer}.
\newblock Design of the chiron high-resolution spectrometer at ctio (2010).
\newblock \doi{10.1117/12.856709}.
\newblock \urlprefix\url{https://doi.org/10.1117/12.856709}

\bibitem{2017ExA....43..167T}
M.~{Tala}, L.~{Vanzi}, G.~{Avila}, C.~{Guirao}, E.~{Pecchioli}, A.~{Zapata},
  F.~{Pieralli}, {Two simple image slicers for high resolution spectroscopy},
  Experimental Astronomy \textbf{43}, 167 (2017).
\newblock \doi{10.1007/s10686-017-9526-5}

\bibitem{2012SPIE.8446E..81B}
A.~{Berdja}, L.~{Vanzi}, A.~{Jord{\'a}n}, S.~{Koshida}, in \emph{Ground-based
  and Airborne Instrumentation for Astronomy IV}, \emph{\procspie}, vol. 8446
  (2012), \emph{\procspie}, vol. 8446, p. 844681.
\newblock \doi{10.1117/12.925572}

\bibitem{1972ailt.conf..227B}
A.~{Baranne}, in \emph{Auxiliary Instrumentation for Large Telescopes}, ed. by
  S.~{Laustsen}, A.~{Reiz} (1972), pp. 227--239

\bibitem{2000ApOpt..39.2614G}
R.G. {Gratton}, R.~{Bhatia}, A.~{Cavazza}, {Asymmetric White-Pupil Collimators
  for High-Resolution Spectrographs}, \ao \textbf{39}, 2614 (2000).
\newblock \doi{10.1364/AO.39.002614}

\bibitem{2009arXiv0910.0167B}
M.A. {Bershady}, {3D Spectroscopic Instrumentation}, ArXiv e-prints  (2009)

\bibitem{1967ApOpt...6.1976S}
D.J. {Schroeder}, {An echelle spectrometer-spectrograph for astronomical use},
  \ao \textbf{6}, 1976 (1967).
\newblock \doi{10.1364/AO.6.001976}

\bibitem{2008arXiv0805.0091L}
D.B. {Leviton}, B.J. {Frey}, {Temperature-dependent absolute refractive index
  measurements of synthetic fused silica}, ArXiv e-prints  (2008)

\bibitem{2018arXiv180407441V}
L.~{Vanzi}, A.~{Zapata}, M.~{Flores}, R.~{Brahm}, M.~{Tala Pinto}, S.~{Rukdee},
  M.~{Jones}, S.~{Ropert}, T.~{Shen}, S.~{Ramirez}, V.~{Suc}, A.~{Jordan},
  N.~{Espinoza}, {Precision stellar radial velocity measurements with FIDEOS at
  the ESO 1-m telescope of La Silla}, ArXiv e-prints  (2018)

\bibitem{2014SPIE.9147E..89T}
M.~{Tala}, A.~{Berdja}, M.~{Jones}, L.~{Vanzi}, S.~{Ropert}, M.~{Flores},
  C.~{Viscasillas}, in \emph{Ground-based and Airborne Instrumentation for
  Astronomy V}, \emph{\procspie}, vol. 9147 (2014), \emph{\procspie}, vol.
  9147, p. 914789.
\newblock \doi{10.1117/12.2056551}

\bibitem{article}
S.~Rukdee, C.~Park, K.M. Kim, S.H. Lee, M.Y. Chun, I.S. Yuk, H.Y. Oh, H.K.
  Jung, L.~chung uk, H.~Lee, M.~D.~Rafal, S.~Barnes, D.~T.~Jaffe, Igrins mirror
  mount design for three off-axis collimators and one slit-viewer fold mirror
  \textbf{29}, 233 (2012)

\bibitem{unknown}
C.~Schwab, N.~Jovanovic, T.~Feger, M.~Bakovic, Y.~V.~Gurevich, J.~Stürmer,
  R.~Apodaca, L.~Vanzi, S.~Rukdee, J.~S.~Lawrence, D.~Coutts, N.~Cvetojevic,
  S.~Mahadevan, G.~K.~Stefánsson, S.~P.~Halverson, O.~Guyon.
\newblock Adaptive optics fed single-mode spectrograph for high-precision
  doppler measurements in the near-infrared (2016)

\bibitem{2015AdSpR..55.2509L}
H.A.N. {Le}, S.~{Pak}, D.T. {Jaffe}, K.~{Kaplan}, J.J. {Lee}, M.~{Im},
  A.~{Seifahrt}, {Exposure time calculator for Immersion Grating Infrared
  Spectrograph: IGRINS}, Advances in Space Research \textbf{55}, 2509 (2015).
\newblock \doi{10.1016/j.asr.2015.03.007}

\bibitem{MNR:MNR21382}
L.~Vanzi, J.~Chacon, K.G. Helminiak, M.~Baffico, T.~Rivinius, S.~Štefl,
  D.~Baade, G.~Avila, C.~Guirao, Pucheros: a cost-effective solution for
  high-resolution spectroscopy with small telescopes, Monthly Notices of the
  Royal Astronomical Society \textbf{424}(4), 2770 (2012).
\newblock \doi{10.1111/j.1365-2966.2012.21382.x}.
\newblock \urlprefix\url{http://dx.doi.org/10.1111/j.1365-2966.2012.21382.x}

\bibitem{1992nstc.rept.....L}
S.D. {Lord}, {A new software tool for computing Earth's atmospheric
  transmission of near- and far-infrared radiation}.
\newblock Tech. rep. (1992)

\bibitem{2000A&A...354.1134R}
P.~{Rousselot}, C.~{Lidman}, J.G. {Cuby}, G.~{Moreels}, G.~{Monnet}, {Night-sky
  spectral atlas of OH emission lines in the near-infrared}, \aap \textbf{354},
  1134 (2000)

\bibitem{2008ApJS..178..374K}
F.~{Kerber}, G.~{Nave}, C.J. {Sansonetti}, {The Spectrum of Th-Ar Hollow
  Cathode Lamps in the 691-5804 nm region: Establishing Wavelength Standards
  for the Calibration of Infrared Spectrographs}, \apjs \textbf{178}, 374-381
  (2008).
\newblock \doi{10.1086/590111}

\bibitem{2011ApJS..195...24R}
S.L. {Redman}, J.E. {Lawler}, G.~{Nave}, L.W. {Ramsey}, S.~{Mahadevan}, {The
  Infrared Spectrum of Uranium Hollow Cathode Lamps from 850 nm to 4000 nm:
  Wavenumbers and Line Identifications from Fourier Transform Spectra}, \apjs
  \textbf{195}, 24 (2011).
\newblock \doi{10.1088/0067-0049/195/2/24}

\bibitem{Whaling2002}
W.~Whaling, W.H.C. Anderson, M.T. Carle, J.W. Brault, H.A. Zarem.
\newblock Argon i lines produced in a hollow cathode source, 332 nm to 5865 nm
  (2002)

\bibitem{opencv_library}
G.~Bradski, {The OpenCV Library}, Dr. Dobb's Journal of Software Tools  (2000)

\bibitem{1983ats..book.....P}
B.A. {Palmer}, R.~{Engleman}, \emph{{Atlas of the Thorium spectrum}} (1983)

\end{thebibliography}

\newpage
\appendix

\section{Sensitivity Analysis}

\begin{figure}[h]
\centering\includegraphics[width=0.7\textwidth]{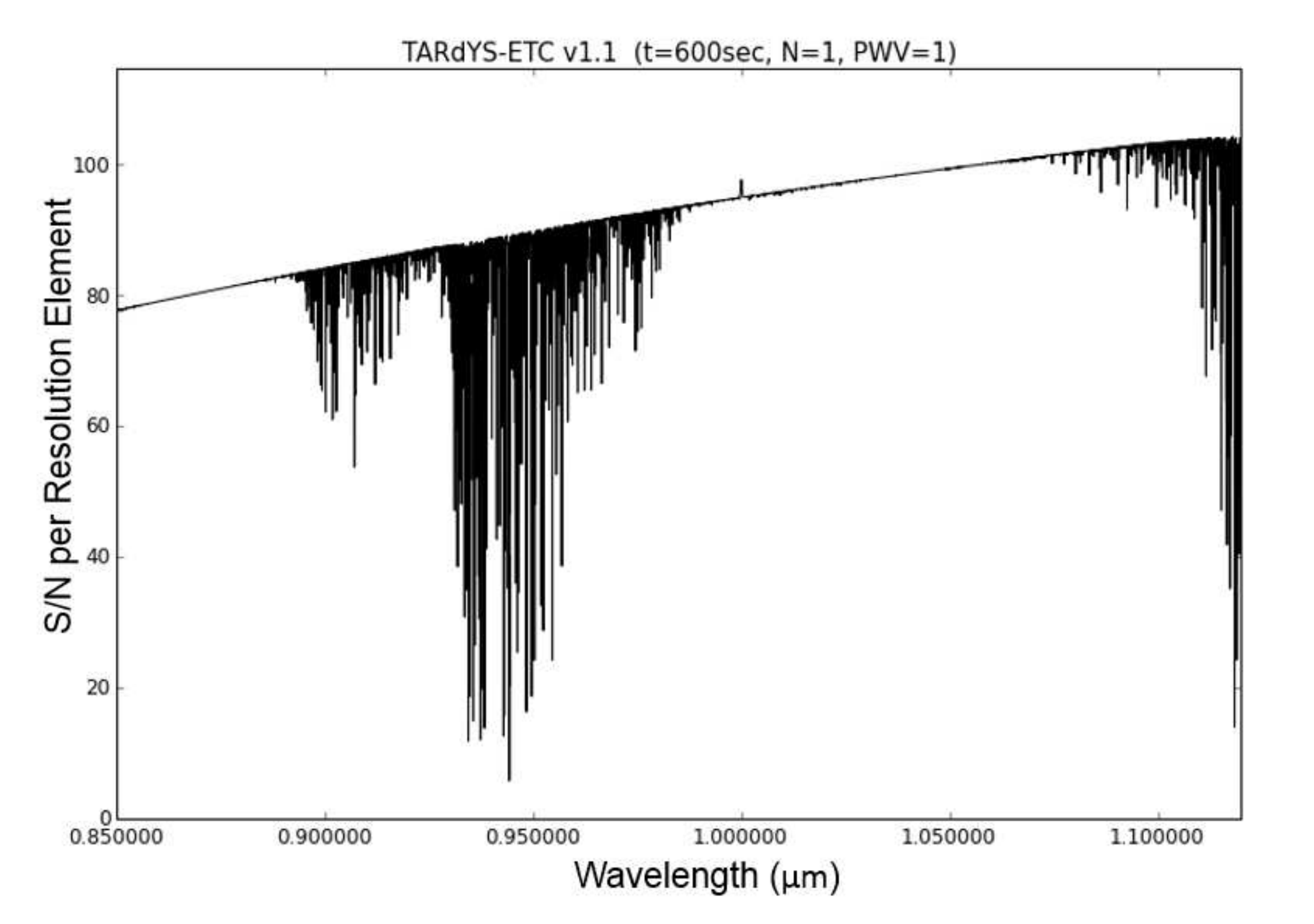}
\caption{Simulation of the signal to noise per resolution element of TARdYS while targeting an M type star. We apply the efficiency budget of the optical components employed in the final optical design of the spectrograph. The black region represents a gap between two photometric bands due to the atmospheric transmission.}
\label{fig:SN}
\end{figure}

We estimate the instrument's performance on the sky in a realistic observing condition. We consider a 3000K M-type star, and  use realistic efficiency estimates. Then we simulate signal to noise per resolution element shown in Fig. \ref{fig:SN}. The detail of simulation method has been presented in \cite{2015AdSpR..55.2509L}. Our throughput parameters used for this simulation are shown in table \ref{tab:e_budget}. In this simulation, we assume the telescope size of 6.5 m, thoughput of 10 \%, readout noise of 5 electrons (for IR array), and dark current of 0.02 electron/s. 

\begin{table}[h!]
\caption {Spectrograph efficiency budget} 
\label{tab:e_budget}
\begin{center}
\begin{tabular}{ lc }
  \hline
  Parameter & Efficiency \\  \hline
  Atmosphere & 0.85 \\
  Telescope and interface & 0.55 \\
  FRD* \cite{MNR:MNR21382} & 0.75 \\
  Doublet (Input) & 0.96 \\
  Collimator & 0.95 \\
  Echelle & 0.55 \\
  VPH & 0.90 \\
  Camera & 0.90 \\
  Cut off filter & 0.95 \\
  Cold window & 0.95 \\
  Detector \cite{2015AdSpR..55.2509L} & 0.80 \\
  \hline
\end{tabular}
\end{center}
\center{*FRD = fiber focal ratio degradation}
\end{table}
In our simulation, we include telluric absorption spectra from ATRAN \cite{1992nstc.rept.....L} for the resolution R = 60,000, which is an expected resolution of this spectrograph. We consider the telescope's location in Atacama desert in Chile and therefore simulate a low precipitable water vapor (PWV) . The telluric OH Emission is obtained from \cite{2000A&A...354.1134R}. To predict the performance of the spectrograph, we simulate the signal to noise (S/N) ratio per resolution element cover the near-infrared Y band with rest wavelength of 1 $\mu$m as shown in Fig. \ref{fig:SN}. We assume initial parameter as follows: PWV = 1 mm, single exposure of 10 minutes, seeing of 1 arcsecond and no moonlight. The blackbody temperature is set to 3000K for M dwarfs observation.

\begin{figure}[h]
\centering\includegraphics[width=0.7\textwidth]{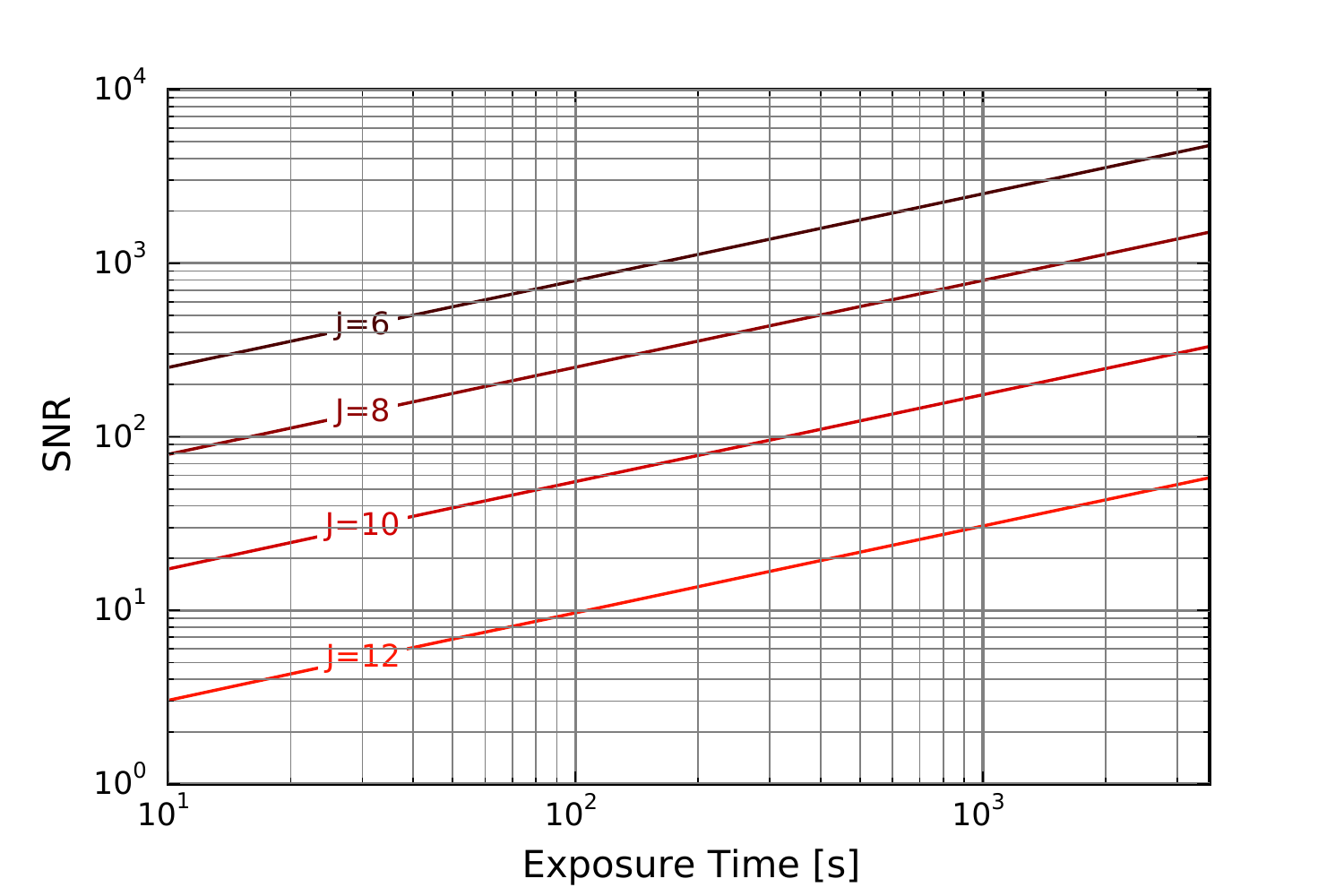}
\caption{The signal to noise prediction for average observing conditions during 30-minute exposure time varies on the x-axis, assuming a system efficiency of 10\%. Stars with different magnitudes are shown. The SNR is calculated for the Y-band magnitude while J6 = Y6.63, J8 = Y8.63, J10 = Y10.63, J12 = Y12.63 respectively. We assume the blackbody temperature to be 3000K.}
\label{fig:maglim}
\end{figure}

In Fig. \ref{fig:maglim}, we present the simulated prediction of signal to noise as a function of exposure time. We display the Y-band signal to noise calculation with the J-magnitude label because it is more common to find the J-magnitude reference the object of interest in the currently available catalogs.

\section{Wavelength Calibration Source}

\begin{figure}[ht!]
\centering\includegraphics[scale=0.5]{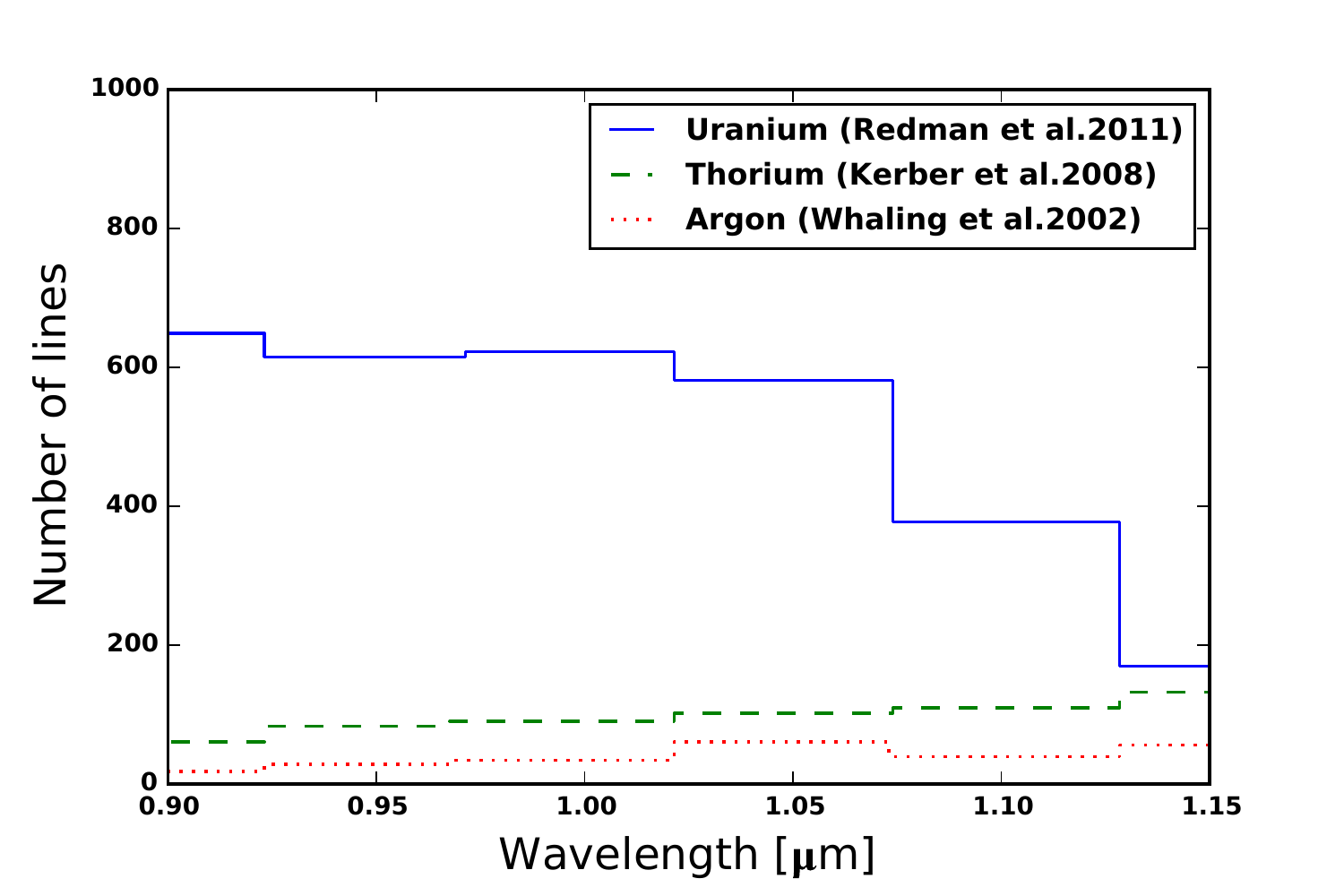}
\caption{Histograms of the number of lines of standard uranium compared with thorium and argon (taken from \cite{2008ApJS..178..374K}; \cite{2011ApJS..195...24R}; \cite{Whaling2002}). Note that the distributions are collected from various sources; they are not necessarily proportional to the intrinsic distribution density.}
\label{fig:calib_line2}
\end{figure}

The wavelength calibration source is worth mentioning in this study because it plays a crucial role in achieving RV precision. In optical astronomy, thorium-argon (ThAr) lamps are widely used as a calibration source. However, in Y-band in the infrared ($\lambda$ = 0.9-1.1 $\mu$m), uranium (U) lamps exhibit more lines. 
In Fig. \ref{fig:calib_line2} we compare the number of lines of ThAr and U lamps in our working wavelength. U lamp yields a much higher number of lines. Thus, we plan to use a uranium lamp instead of the ThAr lamp. The fiber connector receives the image of the calibration source formed by an achromatic doublet lens (f = 100 mm) and feeds the calibration spectrum to the spectrograph fiber link. This produces pairs of spectral orders on the detector that are passed through the same lens, allowing stable, relative wavelength calibration.

\newpage
\section{Data acquisition and reduction}

In the two prototype setups, we use a commercial InGaAs infrared detector XEVA-1.7-640 which is composed of 512x640 pixels of size 20 $\mu$m size for each pixel instead of the H1RG detector because the physical size of XEVA is slightly smaller than that of H1RG. We could not apply the traditional method of data acquisition and reduction. Because of the size difference, we introduce a data reduction and image processing routine to be used specifically for this prototype dataset obtained from XEVA detector. We make a mosaic of the four corners with a relatively large overlapping area. Then we apply a pattern recognition technique to the images in order to merge them together. Each corner consists of a 30 images averaged for a master image. The image is subtracted by a masterdark image. The background level of the XEVA detector is relatively high. To normalize the signal levels, we divide by the background level, obtained from the lowest value of the image.

\begin{figure}[h!]
\centering\includegraphics[scale=0.5]{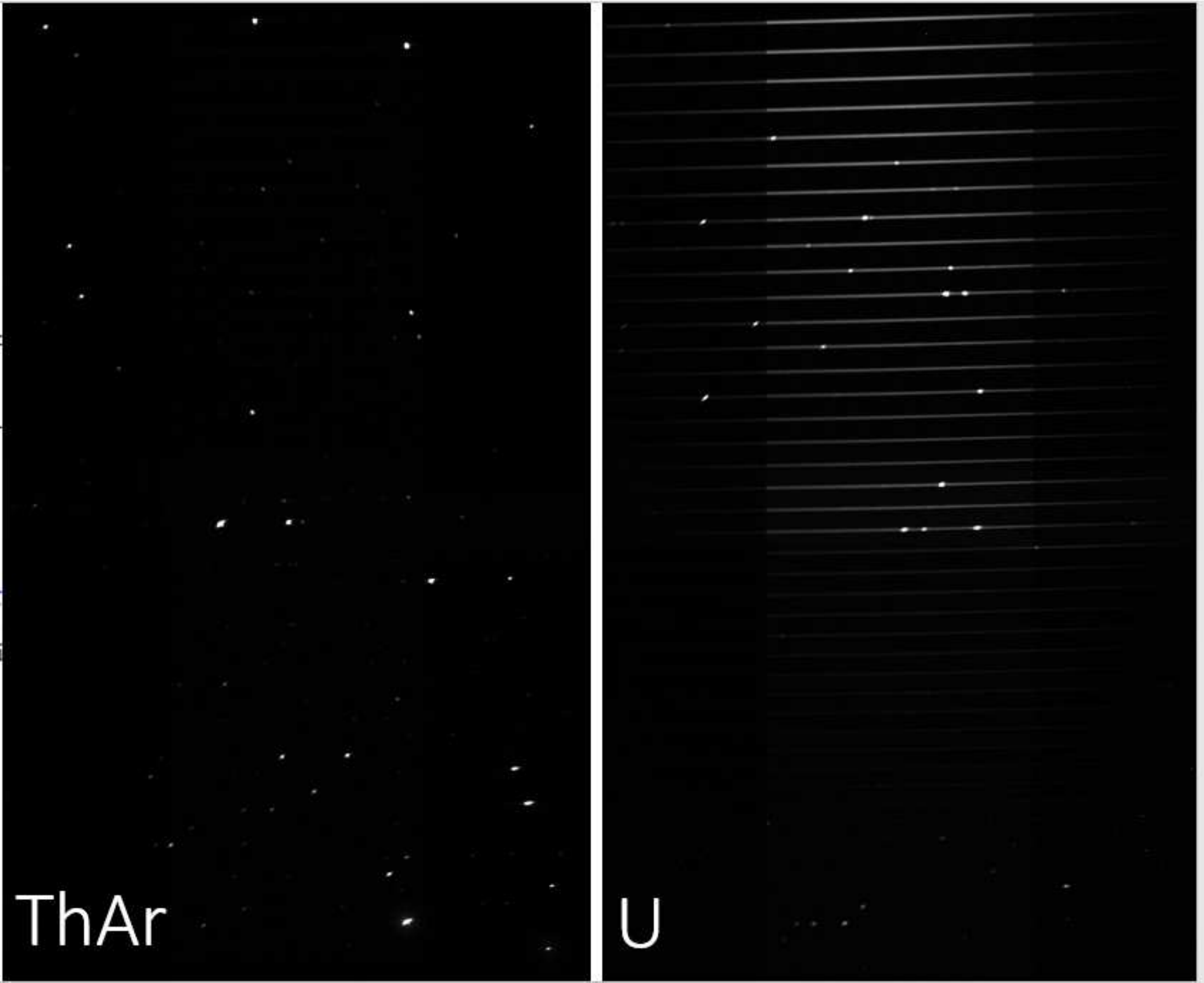}
\caption{The full field merged spectral image obtained from 4 corner positions. The bright region in Uranium (U) spectra is the overlap region. Wavelength orders increase from bottom to top and left to right. The continuum in the orders is likely from the lamp background in the U lamp case.}
\label{fig:calibff}
\end{figure}

\section{Image Processing}

We created a real-time FWHM (Full Width Half Maximum) reader using opencv \cite{opencv_library} apply with a FWHM calculation for each spot. Our 'FWHM analyser' algorithm first converts the images to grayscale, and then automatically calculates the FWHM at each detected point source. One drawback of working in grayscale using opencv is that the image loses resolution intensity. Nevertheless, it is enough to obtain basic information for our prototype. The program determines background levels automatically from the image using the heuristic $(maximum-median)\times 10\% +median$. We perform open and close transforms to remove noisy edges from each spot found with contours from the command cv.findContours, and thereby get a bounding rectangle. Next, any spots lying above this threshold are identified. Finally, the FWHM of each spot is determined by finding the maximum and measuring the radius of the best-fitting circle around pixels above half maximum by fitting a Gaussian and background level simultaneously. 

Separately, we implement a script to merge the XEVA corner images into a full field image from four mosaic patches. First we merge the two upper halves together and then merge the upper halves and lower halves. In the images obtained from the calibration lamps, we only have spots. We implemented an algorithm to detect all spots of the two frames that we want to merge and overlap the two images. The spots detection is very similar to the one detected by FWHM analyser. Our algorithm is robust enough to allow the background adjustment so we make sure that the spots use in the stitch program are from the light source, and are not noise from the detector. We check the size (FWHM) and intensity of all detected spots, calculate the similarity of the matching spots, and find the closes distances of the spots. Then we merge the two images at a time according to matching information. In the end, we obtain a merged image of full field spectra as shown in Fig. \ref{fig:calibff}.

\begin{figure}[h!]
\centering\includegraphics[scale=0.4]{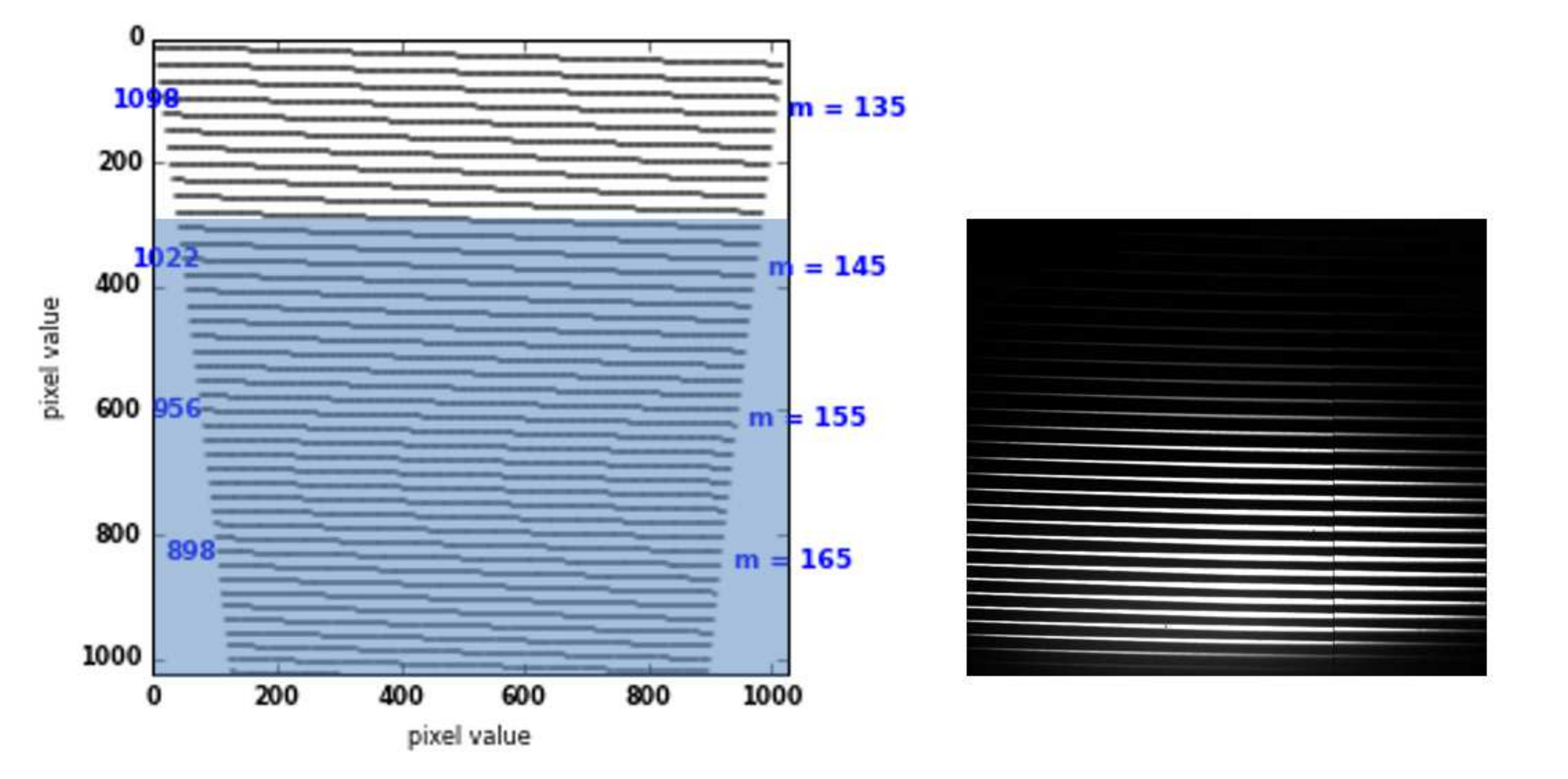}
\caption{Comparison of the simulation (left) and observed continuum spectra (right) obtained from an infrared LED, which has a working wavelength from 0.86-0.98 $\mu$m. The spectra (right) corresponds to the 147th-175th order. as shaded in simulation image in blue(left).}
\label{fig:comparesimureal}
\end{figure}

\begin{figure}[h!]
\centering\includegraphics[scale=0.55]{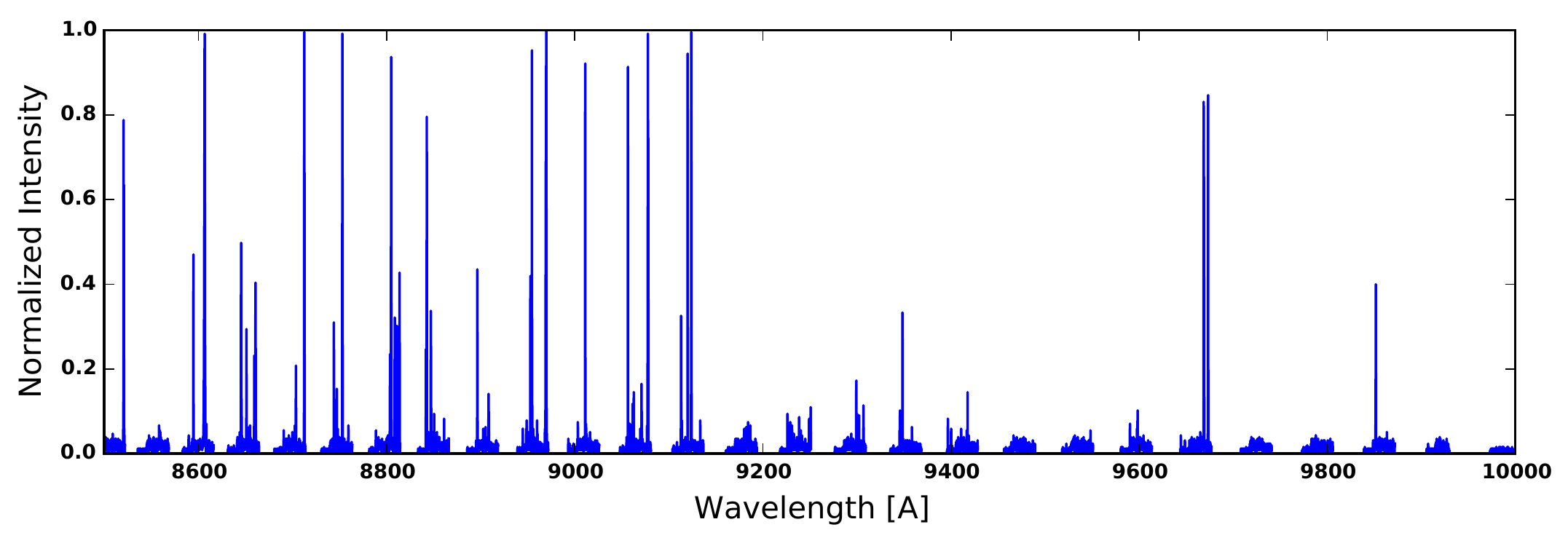}
\caption{Merged spectrum using the IR LED continuum reference on the ThAr calibration image in the range of 0.86-0.98 $\mu$m. Each order is separated by gaps in the plot.}
\label{fig:mergespec}
\end{figure}

To identify the spectral orders, we use a low cost IR LED of approximate wavelength range 0.86-0.98 $\mu$m as our reference. This range covers about two thirds of the spectral range. Fig. \ref{fig:comparesimureal} shows the comparison between the simulated data before. In our simulation, the smallest order separation is 325 $\mu$m, which corresponds to about 16 pixels (of 20 $\mu$m size per each pixel) on the XEVA detector. Each order is extracted using the central wavelength reference and echellogram calculation (Fig. \ref{fig:echellogram}). To obtain more information, we extract with a 3 pixel-wide band. We extract the data from every order (limited to the IR continuum reference) and calibrate it with an IRAF spectral atlas \cite{1983ats..book.....P} as shown in Fig. \ref{fig:mergespec}. 

\newpage
\section{Temperature Management}
\label{sec:tempmanagement}

\begin{figure}[h!]
\centering\includegraphics[scale=0.7]{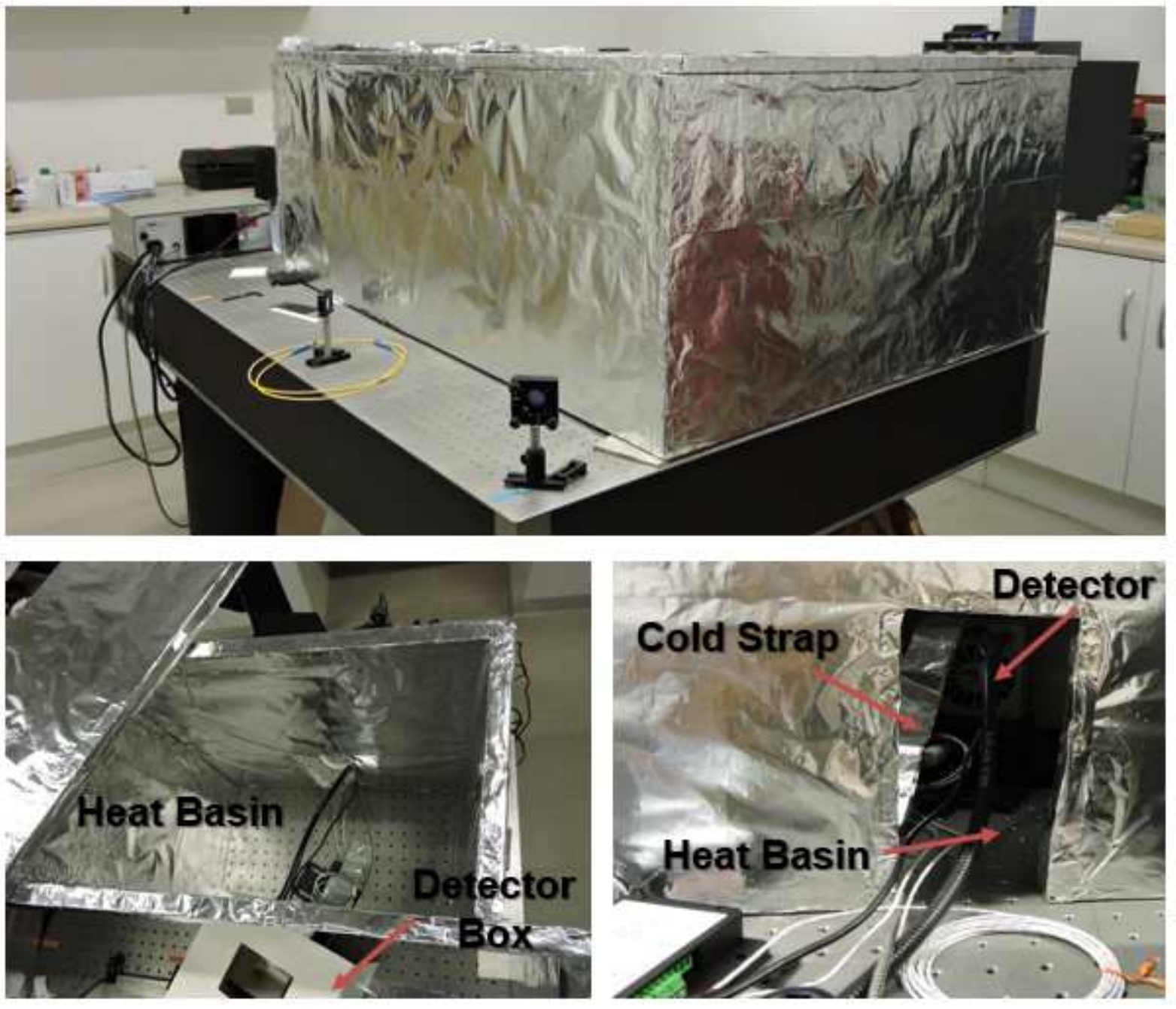}
\caption{The simple implemented enclosure using foam board wrapped with a single layer aluminium foil (top panel). At the back of this enclosure (bottom panel) we attach a heat basin for Experiment 3 which acts as a heat filter between inside and outside of the enclosure through a small opening.}
\label{fig:enclosure}
\end{figure} 

In the experiment, we built a simple enclosure with a heat basin and a thermal-strap on the detector. The main idea of the temperature stability control is to avoid directly exposing the temperature sensor to the air in the room, which is controlled by AC. Instead, we let the air circulate in the system, balancing stability and smoothing room temperature fluctuations. This enclosure in Fig. \ref{fig:enclosure} consists of three different parts: 1. the main enclosure, 2. the heat basin at the back of the detector and 3. the detector enclosure. The main enclosure covers the spectrograph optics, and the basin area acts as an air buffer. The XEVA detector, which is used for the prototype, releases hot air from the back of the detector and cooled air at the side. Normally, the air from the detector will disturb the bench temperature sensor. Thus, we also installed a small box to cover the detector and the last lens together. The material of this box consists of a very thin layer of foam for insulation (see Fig. \ref{fig:enclosure} bottom left). This small box has an opening to the heat basin at the back of the detector. It encloses the side of the detector with another small opening at the top. The opening at the back of the enclosure lets in cool air from the bottom to circulate the heat in the system. Additionally, we install a simple cold strap composed from 100 aluminum foil sheets (Fig. \ref{fig:enclosure} bottom right) connected to the detector box on one end and release the heat to the basin on the other end.

\begin{figure}[h!]
\centering\includegraphics[scale=0.6]{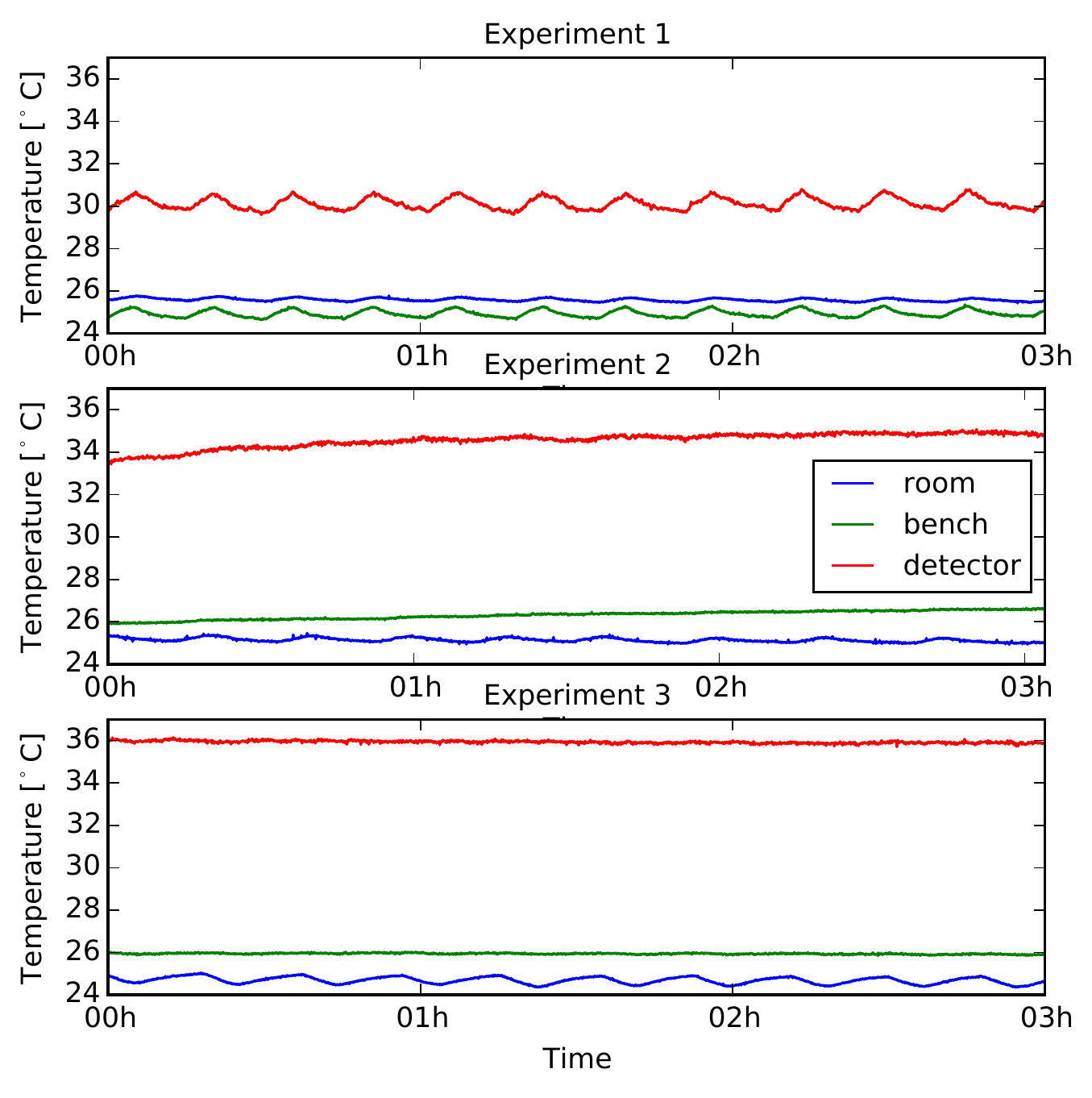}
\caption{Temperature monitoring of the three experiments after the detector reaches a stable temperature. Experiment 1 is the setup without enclousure. Experiment 2 is the setup of a closed system. Experiment 3 is the semi-open system.}
\label{fig:tempcombine}
\end{figure}

We compare the temperature stability of three different experiments in Fig.  \ref{fig:tempcombine}. We place three sensors on (1) a table in the room (blue line), (2) the optical bench (green line), and (3) the detector enclosure (red line). The top panel shows the temperature monitoring from experiment 1, without the enclosure. All sensors fluctuate according to the air-conditioner (AC) cycle approximately every 20 minutes. The detector sensor heats up the most, followed by the room and then the bench. The results from the closed environment in experiment 2 are showing in the middle panel. The temperature sensor on the optical bench and detector is not triggered by the room AC cycle anymore. However, both heat up because of the heat produced by the detector. Over the three hours of the experiment, the bench temperature rose by 0.5 $^\circ$C, and the detector increased 1.0 $^\circ$C, continuously. No stable stage was seen in either. Lastly, in our semi-open system solution (experiment 3), we can control the temperature of the bench and detector such that they remain stable at the $\pm$ 0.1 $^\circ$C level and achieve a thermally stable system.

%
%\subsubsection{Experiment 3: Semi-opened system enclosure}

% Non-BibTeX users please use
%\begin{thebibliography}{}

%
% and use \bibitem to create references. Consult the Instructions
% for authors for reference list style.
%
%\bibitem{RefJ}
% Format for Journal Reference
%Author, Article title, Journal, Volume, page numbers (year)
% Format for books
%\bibitem{RefB}
%Author, Book title, page numbers. Publisher, place (year)
% etc
%\end{thebibliography}

\end{document}